\documentclass[11pt]{article}

\usepackage{amsmath}
\usepackage{amsfonts}
\usepackage{latexsym}
\usepackage{epsfig}
\usepackage{color}
\if@twoside\oddsidemargin25mm
\else
\renewcommand{\d}{\mathrm{d}}

\newcommand{\pp}[2][]{\ensuremath{\frac{\partial #1}{\partial #2}}}

\newcommand{\Ra}{\Rightarrow}

\DeclareMathSymbol{\mg}{\mathrel}{symbols}{"1D}

\newcommand{\ga}{\alpha}
\newcommand{\gb}{\beta}
\renewcommand{\gg}{\gamma}
\newcommand{\gd}{\delta}
\renewcommand{\ge}{\epsilon}

\newcommand{\gf}{\phi}

\newcommand{\gx}{\xi}
\newcommand{\gm}{\mu}
\newcommand{\gn}{\nu}

\newcommand{\gk}{\kappa}
\newcommand{\gl}{\lambda}
\newcommand{\gr}{\rho}
\newcommand{\gth}{\theta}
\newcommand{\gs}{\sigma}

\newcommand{\go}{\omega}
\newcommand{\gz}{\zeta}
\newcommand{\gp}{\pi}
\newcommand{\gps}{\psi}
\newcommand{\get}{\eta}
\newcommand{\gch}{\chi}
\newcommand{\gG}{\Gamma}
\newcommand{\gD}{\Delta}

\newcommand{\gL}{\Lambda}

\newcommand{\gTh}{\Theta}

\newcommand{\cD}{{\cal D}}

\newcommand{\cG}{{\cal G}}

\newcommand{\cO}{{\cal O}}

\newcommand{\ui}{{\underline i}}
\newcommand{\uj}{{\underline j}}
\newcommand{\uk}{{\underline k}}
\newcommand{\ul}{{\underline l}}

\newcommand{\tA}{{\tilde A}}

\newcommand{\tF}{{\tilde F}}

\newcommand{\Tr}{\mbox{Tr}}
\newcommand{\tr}{\text{tr}}

\newcommand{\Id}{\text{\small 1}\hspace{-3.5pt}\text{1}}

\newcommand{\ra}{\rightarrow}

\newcommand{\der}{\partial}

\newcommand{\inv}{^{-1}}
\newcommand{\dsp}{\displaystyle}

\newcommand{\ubar}[1]{{\underline{#1}}}

\newcommand{\labl}[1]{\label{#1}}
\newcommand{\Kh}{K\"{a}hler}
\newcommand{\beq}{\begin{equation}}
\newcommand{\eeq}{\end{equation}}
\newcommand{\barr}{\begin{array}}
\newcommand{\earr}{\end{array}}
\newcommand{\equ}[1]{\begin{gather} #1 \end{gather}}

\newcommand{\tabu}[2]{\begin{tabular}{#1} #2 \end{tabular}}
\newcommand{\arry}[2]{\begin{array}{#1} #2 \end{array}}

\newcommand{\pmtrx}[1]{\begin{pmatrix} #1 \end{pmatrix}}

\newcommand{\non}{\nonumber}
\newcounter{oldcounter}

\newcommand{\fZ}{\mathfrak{ Z}}
\newcommand{\bh}{{\bar h}}

\newcommand{\bq}{{\bar q}}

\newcommand{\bz}{{\bar z}}
\newcommand{\bge}{{\bar\epsilon}}

\newcommand{\bgth}{{\bar\theta}}

\newcommand{\bget}{{\bar\eta}}
\newcommand{\bgch}{{\bar\chi}}
\newcommand{\tgg}{{\tilde \gamma}}

\newcommand{\Intr}{\mathbb{Z}}
\newcommand{\Cplx}{\mathbb{C}}

\newcommand{\ba}[2]{\[\begin{array}{#2}\label{#1}}
\newcommand{\ea}{\end{array}\]}
\newcommand{\be}{\begin{equation}}
\newcommand{\ee}{\end{equation}}
\newcommand{\bea}{\begin{eqnarray}}
\newcommand{\eea}{\end{eqnarray}}

\newcommand{\E}[1]{\mathrm{E_{#1}}}
\newcommand{\U}[1]{\mathrm{U(#1)}}
\newcommand{\SU}[1]{\mathrm{SU(#1)}}
\newcommand{\SO}[1]{\mathrm{SO(#1)}}

\newcommand{\ch}[2]{{\text{ch}_{#1}{[#2]}}}

\newcommand{\rep}[1]{\mathbf{#1}}
\newcommand{\crep}[1]{\overline{\rep{#1}}}

\begin{document}

\begin{flushright}
hep-th/0303101
\\
UVIC-TH/03-03
\\
\end{flushright}
\vskip 2 cm
\begin{center}
{\Large {\bf 
Localized tadpoles of anomalous heterotic $\boldsymbol{\U{1}}$'s 
} 
}
\\[0pt]

\bigskip
\bigskip {\large
{\bf S.\ Groot Nibbelink$^{a,}$\footnote{
{{ {\ {\ {\ E-mail: grootnib@uvic.ca}}}}}}}, 
{\bf H.P.\ Nilles$^{b,}$\footnote{
{{ {\ {\ {\ E-mail: nilles@th.physik.uni-bonn.de}}}}}}},
{\bf M.\ Olechowski$^{c,}$\footnote{
{{ {\ {\ {\ E-mail: Marek.Olechowski@fuw.edu.pl}}}}}}}
{\bf M.G.A.\ Walter$^{b,}$\footnote{
{{ {\ {\ {\ E-mail: walter@th.physik.uni-bonn.de}}}}}}}
\bigskip }\\[0pt]
\vspace{0.23cm}
${}^a$ {\it 
University of Victoria,  Dept.\ of Physics \& Astronomy, \\
PO Box 3055 STN CSC, Victoria, BC, V8W 3P6 Canada.\\
(CITA National Fellow)\\
} 
\vspace{0.23cm}
${}^b$ {\it  
Physikalisches Institut der Universit\"at Bonn, \\
Nussallee 12, 53115 Bonn, Germany.\\
}
\vspace{0.23cm}
${}^c${\it  
Institute of Theoretical Physics, Warsaw University, \\
Ho\.za 69, 00--681 Warsaw, Poland.\\
}
\bigskip
\vspace{1.4cm} 
\end{center}
\subsection*{\centering Abstract}

We investigate the properties of localized anomalous $\U{1}$'s in 
heterotic string theory on the orbifold $T^6/\Intr_3$. We argue that
the local
four dimensional and original ten dimensional Green--Schwarz
mechanisms can be implemented simultaneously, making the theory
manifestly gauge invariant everywhere, in the bulk and at the fixed
points. We compute the shape of the 
Fayet--Iliopoulos tadpoles, and cross check this
derivation for the four dimensional auxiliary fields by a direct
calculation of the tadpoles of the internal gauge fields. Finally we
study some 
resulting consequences for spontaneous symmetry breaking, and derive
the profile of the internal gauge field background over the orbifold.

\newpage

\section{Introduction}
\labl{sc:intro}

The present paper is the follow up investigation of our recent work
\cite{Gmeiner:2002es} on localized anomalies in heterotic orbifold
models. Let us therefore briefly summarize the general context and the
main findings of that article. We considered the effective field
theory description of ten dimensional heterotic string theory compactified on
the six dimensional orbifold $T^6/\Intr_3$. Strings on orbifolds have been
discussed first by the authors of refs.\
\cite{dixon_85,Dixon:1986jc} and with the inclusion of non--trivial
gauge field backgrounds, so--called Wilson lines, in
\cite{ibanez_87,Ibanez:1988pj,Font:1988tp}. Recently there 
has been a strong effort to understand the shape of anomalies on 
orbifolds. First in ref.\ \cite{Arkani-Hamed:2001is} the anomalies on 
$S^1/\Intr_2$ were computed and it was found that they localize
at the fixed points of this orbifold. Afterwards, various groups
computed anomalies on the orbifolds $S^1/\Intr_2$, 
$S^1/\Intr_2 \times \Intr_2'$ and $T^2/\Intr_2$
\cite{Scrucca:2001eb,Barbieri:2002ic,Pilo:2002hu,GrootNibbelink:2002qp,Asaka:2002my}.
These results, and the questions on anomaly cancellation in heterotic
orbifold models raised in \cite{Gmeiner:2002ab}, led us to
calculate the gaugino anomaly in ref.\  \cite{Gmeiner:2002es}.
The following two observations form the main conclusions of that
investigation:  
\begin{enumerate}
\item 
First of all, the untwisted bulk gaugino states lead to localized
anomalies at the fixed points of $T^6/\Intr_3$. These anomalies are
entirely determined by the local spectra of those untwisted states, that
survive the orbifold projections at the corresponding fixed points. By taking 
the twisted states at the fixed points into account, we showed that
no non--Abelian anomalies arise at any of the fixed points. 
\item 
However, the structure of the localized anomalous $\U{1}$'s turned out
to be more complicated. Using the fact that the spectrum of a model
with Wilson lines at each fixed point is equivalent to the spectrum
of a model without Wilson lines, it followed, that at most
one of essentially two types of anomalous $\U{1}$'s can be present
locally at each fixed point. The sum of the local anomalous $\U{1}$ generators
corresponds to the possible anomalous $\U{1}$ generator of the zero
mode theory. If this sum vanishes, no anomalous $\U{1}$ appears at the
zero mode level. 
\end{enumerate}

The appearance of global anomalous $\U{1}$'s in heterotic orbifold
compactifications has been
studied extensively in the past and we would like to remind the reader of
the most important results (see ref.\ \cite{Dine:1987xk} for details). 
In heterotic models at most one anomalous $\U{1}$ exists at the zero
mode level. 
Gauge invariance is restored by a four dimensional
remnant of the Green--Schwarz mechanism \cite{Green:1984sg}, which 
leads to the coupling of the model independent axion to the 
anomalous Abelian gauge field 
\cite{Kim:1988dd,Chun:1992xm,Georgi:1998au}. However, as observed in 
\cite{Dine:1987xk}, the sum of the charges does not vanish for the
anomalous $\U{1}$, and therefore a quadratically divergent
Fayet--Iliopoulos (FI) tadpole arises at one-loop \cite{Fischler:1981zk}.
By direct calculations \cite{Atick:1987gy,Dine:1987gj}
of scalar masses it has been confirmed, that this tadpole
arises in string theory as well.\footnote{Similar tadpoles in open
string theory turn out to vanish \cite{Poppitz:1998dj}.} 
However, in that case the string scale $M_s$ provides the cut--off for
the quadratic divergence. In $N=1$ supersymmetric field
theories in four dimensions, the Fayet--Iliopoulos D--term can either
lead to supersymmetry or gauge symmetry breaking \cite{Fayet:1974jb}. 
For heterotic orbifold models only the latter possibility seems to be
realized: the anomalous $\U{1}$ is spontaneously broken; its gauge
field acquires a mass of the order of the string scale, which
effectively removes it from the low--energy spectrum.

With these introductory remarks in mind, we are now in the position to
formulate the central issues we wish to address in this work. 
Comparing the situation of the zero mode anomalous $\U{1}$ in
heterotic orbifold models to the structure of localized 
anomalous $\U{1}$'s at the orbifold fixed points, the
following questions naturally arise: 
\begin{itemize}
\item 
How is local gauge invariance restored at the fixed points of $T^6/\Intr_3$? 
\item 
What is the profile of the Fayet--Iliopoulos tadpoles over this orbifold?
\item
What are the consequences of these tadpoles? 
\end{itemize}
As for the first question, we will show, that by a local version of the
four dimensional Green--Schwarz mechanism the local Abelian anomalies
are canceled at the various fixed points.

The structure of Fayet--Iliopoulos tadpoles on orbifolds has received
a lot of attention recently. The existence of quadratically divergent
tadpoles on five dimensional supersymmetric orbifolds, like $S^1/\Intr_2$ and
$S^1/\Intr_2 \times \Intr_2'$, was realized in
\cite{Ghilencea:2001bw} and the shapes of these tadpoles over such
orbifolds have been computed in refs.\  
\cite{Barbieri:2001cz,Scrucca:2001eb,Barbieri:2002ic}. These tadpoles 
of the even auxiliary field components of the five dimensional gauge
super multiplets  
possess both quadratically and logarithmically divergent parts. The latter
are proportional to the double derivative of the fixed point delta
functions. As noticed in ref.\ \cite{Mirabelli:1998aj}, at these branes
the auxiliary field of the four dimensional gauge multiplet is shifted
by the derivative of the odd real scalar of the gauge
multiplet. Therefore it has the same tadpole structure as the even 
auxilary field component. In refs.\ 
\cite{GrootNibbelink:2002wv,GrootNibbelink:2002qp} it was shown,
that such localized tadpoles lead to peculiar shapes of the
corresponding real scalar background, which, in turn, often gives rise
to strong localization of bulk
states to one or both fixed points. This effect appears in particular
due to the double derivatives of the fixed point delta functions. 
For gauge theories in six dimensions compactified on two
dimensional orbifolds, like $T^2/\Intr_2$ and $T^2/\Intr_4$, tadpoles
were found for the internal part of the gauge field strength $F_{56}$ 
at the fixed points \cite{vonGersdorff:2002us,Csaki:2002ur}.

With these results in mind, one can speculate on the properties of tadpoles
in heterotic models on $T^6/\Intr_3$. One complication is, that the ten
dimensional super Yang--Mills theory is only known on--shell. Therefore, 
one cannot directly identify the Fayet--Iliopoulos tadpoles. However,
with respect to the remaining $N=1$ supersymmetry in four dimensions,
one may introduce the appropriate   
auxiliary fields by hand. In addition, as mentioned above, for the
anomalous $\U{1}$'s one
may expect tadpoles of the internal gauge field strengths at the fixed
points. In this paper we introduce
such a four dimensional off--shell formulation, and explicitly compute
the tadpoles of the corresponding auxiliary components and the internal gauge
fields. The comparison of the results for these two types of tadpoles
provides an important cross check of our computations. Motivated by
the results in five dimensions, we also investigate some consequences of these
localized tadpoles.

\subsection*{Paper organization}

In section \ref{sc:HetMod} we introduce the basic elements of
heterotic $\E{8}\times \E{8}'$ theory on the orbifold $T^6/\Intr_3$
with Wilson lines. We explain how the four dimensional $N=1$
supersymmetry, which survives the $\Intr_3$ orbifolding, can be
realized off--shell in the full ten dimensional theory. (The necessary
spinor algebra is reviewed in appendix \ref{sc:Spinor6D}.) This
off--shell formulation makes the coupling of the twisted string
multiplets at the fixed points straightforward. A review of the
possible fixed point equivalent models, that contain (anomalous)
$\U{1}$'s, concludes this section. 

Section \ref{sec:GreenSchwarz} is devoted to the question how the
Green--Schwarz mechanism is realized on the orbifold, such that the local
Abelian anomalies are canceled at the fixed points. Important
factorization properties and trace relations needed to check that our
modifications of the Green--Schwarz action cancel these anomalies, are
provided in appendices \ref{sc:AnomPolyFact} and \ref{sc:E8traces},
respectively.  

Section \ref{sc:Tadpoles}  is devoted to the computation of tadpoles. The
Fayet--Iliopoulos tadpoles, corresponding to the auxiliary $\cD^I$
fields introduced in subsection \ref{sc:EffSYM}, are computed in 
subsection \ref{sc:FItad}. (Properties of wave functions on the torus
$T^6$ can be found in appendix \ref{sc:WaveProp}). To confirm these
results, we calculate the tadpoles for the internal gauge fields in
the following subsection. 

In section \ref{sc:Dterms} we 
investigate the consequences of the modifications of the BPS
background equations. The question of spontaneous symmetry breaking is
addressed, and we show that the internal Cartan gauge fields in
general have non--trivial profiles over the orbifold $T^6/\Intr_3$. 

Finally, we conclude with a summary of the main results and give an
outlook on possible further research directions.

\section{Heterotic $\boldsymbol{\Intr_3}$ models with anomalous
$\boldsymbol{\U{1}}$'s} 
\labl{sc:HetMod}

\subsection{Heterotic $\boldsymbol{\E{8}\times \E{8}'}$ supergravity on
$\boldsymbol{T^6/\Intr_3}$}
\labl{sc:HetOrbi}

The low energy description of heterotic $\E{8}\times \E{8}'$ string
theory consists of ten dimensional $N=1$ supergravity coupled to the
super Yang--Mills gauge theory of this group. (For a textbook
introduction see \cite{gsw_2,pol_2}.)  The supergravity multiplet
contains the vielbein $e_M^a$, the dilaton $\gf$, the anti--symmetric
2--tensor $B_{MN}$, the left--handed gravitino $\gps_M$,  and the
right--handed dilatino $\gl$. (Here $M, N$ are ten dimensional
spacetime indices, and $a$ is a corresponding tangent space index.)
The super Yang--Mills theory consists of a ten dimensional gauge field
$A_M$ and a left--handed gaugino $\gch$.  Their adjoint indices 
$\ga=(I, w)$ correspond to the generators  $T_\ga = (H_I, E_{w})$ and
are often repressed for notational simplicity. Here $H_I$ represent the
generators of the Cartan subalgebra, and $E_{w}$ the remaining
generators of $\E{8}\times \E{8}'$ labeled by the root vectors $w$. 
Their components are given by the structure constants in the
Cartan--Weyl basis: $[H_I,E_w]=w_I E_w$. We introduce the notation
$[T_\ga, T_\gb] = f_{\ga\gb}{}^\gg T_\gg$,  
$\tr T_\ga T_\gb = \get_{\ga\gb}$, and 
$\tr [T_\ga, T_\gb] T_\gg = f_{\ga\gb\gg}$. 
Notice that this implies that $f_{Iw}{}^{w'} = w_I \gd_w{}^{w'}$. 
Furthermore we assume that the algebra is normalized such that
$\get_{w w'} = \tr E_w E_{w'} = \gd_{-w\, w'}$.

This theory can be compactified on an orbifold $T^6/\Intr_3$, which
is constructed as follows: The torus $T^6 = \Cplx^3/\gG$ is obtained
from the complex three plane, parameterized by complex coordinates
$z_i$, by modding out the lattice $\gG$,
generated by the identifications $z_i \sim z_i + R_i$ and 
$z_i \sim z_i + \gth_i R_i$. (For the definition of complex
coordinates and their conjugates, $\bz_\ui$, $i,\ui=1,2,3$, in
terms of real coordinates, see \eqref{CmplxBss} in 
appendix \ref{sc:Spinor6D}).
Here $R_i$ denote three real radii of the torus, and 
$\gth_i = \exp(2\pi i\, \gf_i)$ are third roots of unity: 
$3 \gf_i \equiv 0$. (The equivalence relation $a \equiv b$ means that 
$a = b \mod 1$.) The orbifold twist $\gTh$ acts component wise on the
coordinates of the torus $T^{6}$ as 
\(
\gTh ( z_i)  = \gth_i\,  z_i.
\) 
Modding out this twist defines the orbifold $T^6/\Intr_3$. 
From now on we make the convenient choice  
$\gf_i = \frac 13(1^2, \mbox{-}2)$. Then these third roots of unity
are equal $\gth_i = \gth = \exp(2\pi i/3)$. (Notice that 
$\gth +\bgth = -1$, where $\bgth = \gth\inv = \gth^2$ is the
complex conjugate of $\gth$.) This orbifold twist does
not act freely, and therefore results in orbifold fixed points. 
In each of the three complex tori we have three fixed points: 
$\gz_0 = 0,  \gz_1 = \frac 13(2 + \gth)$ and 
$\gz_2 = \frac 13(1 + 2 \gth)$. They are fixed points using shifts over
the lattice of the torus: 
\equ{
 \gth \gz_0 = \gz_0, 
\qquad 
\gth \gz_1 = \gz_1 - 1, 
\qquad 
\gth \gz_2 = \gz_2 - 1 - \gth.
\labl{fixedpoints}
}
Collectively, the $27$ fixed points are denoted by  
\(
\fZ_s = \fZ_{s_1 s_2 s_3} = 
( R_1 \gz_{s_1}, R_2 \gz_{s_2}, R_3 \gz_{s_3} ) 
\)
with the integers $s_1, s_2, s_3 = 0,1,2$.

Since gauge fields are only defined up to group
transformations, the 1--form gauge potential $A_1 = A_M \d x^M$ is not
necessarily invariant under the torus periodicities and the
orbifolding twist. This leads to the introduction of the Wilson lines
$a_j$ ($j = 1,2,3$) and the gauge shift vector $v$ by 
\equ{
\arry{cc}{
A_1(z + \hat\jmath ) = A_1(z + \gth\, \hat\jmath) = T_j A_1(z) T_j\inv,
\quad & \quad  T_j = e^{2 \pi i\, a_j^I H_I}, 
\\[1ex]
A_1(\gTh z) = U A_1(z) U\inv, 
\quad & \quad 
U = e^{2\pi i\, v^I H_I},  
}
}
with $\forall w:~ 3 a_j^I w_I \equiv 0$ and 
$\forall w:~ 3 v^I w_I \equiv 0$.
Here, $\hat\jmath$ and $\gth\, \hat \jmath$ denote the generators of
the torus lattice. The three vectors $\hat\jmath$ have length $R_j$
and are mutually orthogonal. This is the Hosotani mechanism
\cite{Hosotani:1989bm} which  implements the Scherk--Schwarz boundary 
conditions \cite{Scherk:1979ta} for the gauge symmetries.
By combining these conditions with the relations \eqref{fixedpoints},
it is not hard to show, that the following four dimensional untwisted
states 
\equ{
\arry{cc}{
\arry{ll}{
A_{\ui}^{\,~\rep{R}_s}: & 
\rep{R}_s = \{ w ~|~  v_s^I w_I + \mbox{$\frac 13$} \equiv 0 \}, 
\\[1ex]
A_{i}^{~\, \crep{R}_s}: & 
\crep{R}_s = \{ w ~|~ v_s^I w_I +\mbox{$\frac 23$} \equiv 0 \},
}
& 
\arry{ll}{
A_\gm^{~\, \rep{Ad}_s}: & 
\rep{Ad}_s = \left\{ \arry{l}{ \{ I \in \text{Cartan} \}, \\[1ex] 
\{ w ~|~  v_s^I w_I  \equiv 0 \}.} \right. 
}
}
\labl{LocReps}
}
survive the orbifold projection at fixed point $\fZ_s$. Here the local
shift vector $v^I_s = v^I + s_j a_j^I$ is introduced. The gauge group 
corresponding to $\rep{Ad}_s$ is denoted as $G_s$. It is important to
note, that the local shift vectors $v_s$ of all fixed points together
decide whether a consistent string model corresponding
to the gauge shift $v$ and the Wilson lines $a_j$ exists: Modular
invariance requires that the level matching conditions are satisfied 
\equ{
\forall s: ~~ \frac 32 v_s^2 \equiv 0. 
\labl{LevelMatching}
}

We close this subsection with a few words concerning the conventions,
we employ in the remainder of this work.  As
$\smash{A_{i}^{~\,\crep{R}_s}}$ is conjugated to  
$\smash{A_{\ui}^{\,~\rep{R}_s}}$, we may take the latter as
fundamental. (We will see in the next subsection that the $A_\ui$ become  
the $N=1$ supersymmetric partners of left--handed fermions in chiral
multiplets.) From the four dimensional point of view at fixed point
$\fZ_s$ the states $\smash{A_{\ui}^{\,~\rep{R}_s}}$ can 
be interpreted as scalar matter fields in the representation
$(\rep{3}_H,\rep{R}_s)$. The representation $\rep{3}_H$ is with
respect to the holonomy group $\SU{3}_H \subset \SO{6} \subset
\SO{1,9}$. (To be 
precise, the holonomy group of the blow up, the holonomy of the
orbifold is $\Intr_3$.)  Finally, unless otherwise
stated, expressions like $F_{i \ui}$ implicitly assume, that Einstein's
summation convention is employed.

\subsection{Effective four dimensional supersymmetry} 
\labl{sc:4Dsusy}

The $\Intr_3$ orbifold twist is chosen such that only $N=1$
supersymmetry in four dimensions is preserved at the zero mode
level. The twist acts on a six dimensional internal spinor as 
\equ{
\gTh: ~ \get_{\gk_3\gk_2\gk_1} \ra 
e^{ -\gp i \, \gf^i \gk_i} \,  \get_{\gk_3\gk_2\gk_1},
}
where $\gk_i = \pm$ represent the internal two dimensional
chiralities. (Conventions and properties of the six dimensional
spinors used in this work have been collected in appendix
\ref{sc:Spinor6D}.) The components of the original 10--dimensional
supersymmetry parameter $\ge_{10}$, corresponding to the supersymmetry 
which remains unbroken by the orbifolding, can be represented as   
\equ{
\ge_{4} = \get_{+++}\otimes \ge_L - \get_{---} \otimes \ge_R, 
\qquad
\ge^{C_-} = \ge\,.
\labl{4Dsusy}
}
Here $\ge$ is a Majorana spinor w.r.t.\ the four dimensional charge
conjugation matrix $C_-$ and the subscript $L, R$ refers to the
four dimensional chirality. This expression is obtained by
exploiting Majorana--Weyl condition of the supersymmetry parameter
$\ge_{10}$ in ten dimensions (see
\eqref{Expn10DF} of appendix \ref{sc:Spinor6D}). 
Notice, that this decomposition can be applied, even
if $\ge = \ge(x, z)$ is a function of both the four dimensional
Minkowski and orbifold coordinates, $x$ and $z$ respectively.

Following the method of ref.\ \cite{Mirabelli:1998aj} we can decompose
the ten dimensional supersymmetry transformation in terms of the
unbroken four dimensional supersymmetry. Contrary to the five
dimensional situation under investigation in ref.\
\cite{Mirabelli:1998aj}, for the ten dimensional theory there
is no off--shell formulation available. However, by rewriting the ten
dimensional super Yang--Mills such that only the remaining four
dimensional supersymmetry is manifest, it becomes rather
straightforward to infer the $N=1$ four dimensional off--shell
formulation of the theory. As we will see, this approach is particularly
useful to describe the interactions with the twisted states (see
section \ref{sc:InclTwisted}).  

A ten dimensional supersymmetry variation $\ge_{10}$ of the gauge
field $A_M$ and the gaugino $\gch$ read 
\equ{
\gd A_M = \frac 12 \bge_{10} \gG_M \gch, 
\qquad 
\gd \gch = - \frac 14 F^{MN} \gG_{MN} \ge_{10} + \ldots,
}
where the dots represent terms of higher order in the 
fields\footnote{Notice
that we have absorbed a dilaton factor into the definition of the
gaugino as compared to \cite{gsw_2}.
This means that the normalization of
the gauge field and the gaugino is the same; the modifications in the
supersymmetry transformation are higher order in the fermion fields.}, and the field strength is defined as 
\equ{
i F_{MN} = \der_M i A_N - \der_N i A_M + [ iA_M, iA_N ].
} 
By substituting $\ge_{10} = \ge_4$ given in \eqref{4Dsusy} and the
decomposition of the gaugino  \eqref{Expn10DF} of appendix 
\ref{sc:Spinor6D}, and 
using table \ref{tab:CmplxClifford} together with the multiplication rules 
\eqref{BasicCliffordAction} of appendix \ref{sc:Spinor6D}, 
we find the following four dimensional $N=1$ supersymmetry transformations
\equ{
\gd A_\gm = 
\frac 12 \bge_L \gg_\gm \gch^{+++}_L 
+ \frac 12 \bge_R \gg_\gm \gch^{+++}_R, 
\qquad 
\gd \gch^{+++}_L = 
- \frac 14 F^{\gm\gn} \gg_{\gm\gn}\ge_L 
- \frac 12 F_{\ui i} \ge_L 
\labl{4DgauginoSusy}
}
and  
\equ{
\arry{l}{
\gd A_{\ubar 1} =  \frac 12 \sqrt 2 \, \bge_R \gch^{--+}_L, 
\\[1mm]
\gd A_{\ubar 2} =  \frac 12 \sqrt 2 \, \bge_R \gch^{-+-}_L, 
\\[1mm]
\gd A_{\ubar 3} =  \frac 12 \sqrt 2 \, \bge_R \gch^{+--}_L, 
}
\qquad 
\arry{l}{
\gd \gch^{--+}_L  =  
\frac 12 \sqrt 2\, F^{\gm}_{~~\ubar 1} \gg_{\gm}\ge_R 
+ \frac 12 F_{23} \ge_L, 
 \\[1mm]
\gd \gch^{-+-}_L  = 
\frac 12 \sqrt 2\, F^{\gm}_{~~\ubar 2} \gg_{\gm}\ge_R 
+ \frac 12 F_{31} \ge_L, 
\\[1mm]
\gd \gch^{+--}_L  = 
 \frac 12 \sqrt 2\, F^{\gm}_{~~\ubar 3} \gg_{\gm}\ge_R 
+ \frac 12 F_{12} \ge_L.
}
\labl{4DchiralSusy}
}
Using the linear part of the supersymmetry variation of the fermion
the off--shell multiplet structure can be (re)constructed. For a
vector multiplet $(B_\gm, \gr, D)$ and a chiral multiplet 
$(Z, \gz_L, f)$ the supersymmetry transformations read 
\equ{
\arry{l}{
\gd B_\gm = \frac 12 \bge_L \gg_\gm \gr_L 
+ \frac 12 \bge_R \gg_\gm \gr_R,
\\[1ex]
\gd \gr_L = - 
 \frac 14 F^{\gm\gn} \gg_{\gm\gn} \ge_L -  \frac i2 \cD \ge_L, 
}
\qquad 
\arry{l}{
\gd Z = \frac 12 \sqrt 2\, \bge_R \gz_L, 
\\[1ex]
\gd \gz_L = \frac 12 \sqrt 2\, \gg^\gm D_\gm Z \ge_R 
+ \frac 12 \sqrt{2}\,  f \ge_L.
\labl{4DsusyTrans}
}
}
(Taken from \cite{VanProeyen:1983wk}, with 
$\gz \ra \gz/\sqrt{2}$ and some sign changes.) 
Comparing this with the result we obtained above, we can 
read off the multiplet structures and the equations of motion of the
auxiliary fields:
\equ{
\arry{cccc}{
\Bigl( A_\gm, \gch^{+++}, D  \Bigr),
~ & ~
\Bigl( A_{\ubar 1}, \gch^{--+}_L, f_{\ubar 1} \Bigr),
~ & ~
\Bigl( A_{\ubar 2}, \gch^{-+-}_L, f_{\ubar 2} \Bigr),
~ & ~
\Bigl( A_{\ubar 3}, \gch^{+--}_L, f_{\ubar 3}\Bigr),
\\[2ex] 
\cD = i F_{i \ui},
& 
f_{\ubar 1} = \frac 12 \sqrt 2\, F_{23}, & 
f_{\ubar 2} = \frac 12 \sqrt 2\, F_{31}, & 
f_{\ubar 3} = \frac 12 \sqrt 2\, F_{12}. 
}
\labl{BulkMultiplets}
}
As these are ordinary $N=1$ off--shell multiplets in four dimensions,
the standard multiplet calculus, see for example 
\cite{VanProeyen:1983wk,Cremmer:1983en,deWit:1983tq}, 
or superspace methods \cite{wundb}, can be applied. This holds
true even though all these fields are still functions of the internal
dimensions. Alternatively, we could perform Fourier decompositions of
the internal dimensions, but then one has to keep track of many 
Kaluza--Klein towers. Of course, both approaches are equivalent, but
in order to avoid writing complicated sums and to be able to trace local
effects easily, we choose to work in coordinate space.

\subsection{Elements of the super Yang--Mills Action}
\labl{sc:EffSYM}

The ten dimensional Yang--Mills action takes the form 
\equ{
L_{YM} = - \frac 14 \tr F_{MN} F^{MN} = 
- \frac 14 \tr F_{\gm\gn} F^{\gm\gn} 
- \tr F_{\gm i} F^{\gm}_{~\, \ui} 
- \frac 12 \tr F_{i j} F_{\ui \uj} 
- \frac 12 \tr F_{i \uj} F_{\ui j}, 
\labl{DecompYM}
}
in the decomposition to four dimensions. (We have made
the simplifying assumption that the dilaton is constant.) It is instructive to
interpret this action from a four dimensional point of view. The first
term in this equation represents the four dimensional gauge field
Lagrangian. The second term gives the kinetic action for the four
dimensional scalars $A_\ui$: 
\equ{
- \tr F_{\gm i} F^{\gm}_{~\, \ui} = 
- \tr D_\gm A_i D^\gm A_{\ui} - \tr \der_i A_\gm \der_\ui A^\gm 
+ \tr \der_i A_\gm D^\gm A_\ui + \tr \der_\ui A_\gm D^\gm A_i,
} 
with the covariant derivative 
$D^\gm A_\ui = \der^\gm A_\ui + i [A^\gm, A_\ui]$. 
The second term in this expression corresponds to Kaluza--Klein masses
if one would choose to work in momentum space. The last two terms 
constitute the mixing between the massive Kaluza--Klein
excitations of $A_\gm$ and $A_\ui$. 
The final two terms in \eqref{DecompYM} can be expressed as 
\equ{
- \frac 12 \tr F_{i j} F_{\ui \uj} - \frac 12 \tr F_{i \uj} F_{\ui j}
= \non 
- \frac 12 \tr [A_i, A_\ui] [A_j, A_\uj] 
+ i \tr (\der_i A_\ui - \der_\ui A_i) [A_j, A_\uj]
\\
- \tr [A_i, A_j] [A_\uj, A_\ui] 
-i \tr (\der_i A_j - \der_j A_i) [A_\ui, A_\uj] 
-i \tr (\der_\ui A_\uj - \der_\uj A_\ui) [A_i, A_j] 
\labl{6DYMpre}
\\ 
- \tr \der_\uj A_i \der_j A_\ui + 
\frac 12 \tr \der_i A_\uj \der_j A_\ui + 
\frac 12 \tr \der_\ui A_j \der_\uj A_i 
- \frac 12 \tr (\der_i A_j - \der_j A_i) (\der_\ui A_\uj - \der_\uj A_\ui). 
\non 
}
Here we have used the Jacobi identity to rewrite 
$\tr [A_i, A_\uj] [A_\ui, A_j]$, and applied partial integrations to
the term $i\tr \der_\ui A_j [A_\uj, A_i]$ and its conjugate. 
(In ref.\ \cite{Witten:1985xb} it was first realized that by using the
Jacobi identity, the dimensional reduced heterotic theory could be
formulated as $N=1$ supergravity in four dimensions.) Clearly
the first line resembles the structure of $D$--terms in $N=1$
supersymmetry, while the second line takes the form of an $F$--term
potential. To make this four dimensional off--shell structure explicit, 
we use the auxiliary fields $f_i$, $f_{\ui}$ and $\cD$, which were
introduced in
\eqref{BulkMultiplets}, and rewrite that part of the action as 
\equ{
- \frac 12 \tr F_{i j} F_{\ui \uj} - \frac 12 \tr F_{i \uj} F_{\ui j}
= 
\tr \Bigl( 
f_i f_\ui - \frac 12 \sqrt2\, \ge_{ijk} f_\ui F_{\uj\, \uk} 
- \frac 12 \sqrt 2\, \ge_{\ui\, \uj\, \uk} f_i F_{j k} \Bigr) 
+ \frac 12 \tr \cD^2 - i \tr \cD F_{i \ui}. 
\labl{6DYM}
}
In this work we do not attempt to obtain the full superpotential and \Kh\
potential of the heterotic theory, which is fully equivalent to the
original ten dimensional description.

Finally we give some other parts of the super Yang--Mills action we need in
the calculation of tadpoles below. The gaugino Lagrangian is given by 
\equ{
L_{gaugino} = - \frac 12 \bgch \gG^M D_M \gch. 
}
This fermionic action can also be decomposed into four dimensional
fields. However, for the tadpole calculations we present
later, it is more convenient to keep the ten dimensional
structure manifest. 
As usual computations of loop corrections involving gauge fields,
require a gauge fixing prescription in order to be able to define
their propagators. All loop computations in this work are performed
using the ten dimensional Feynman gauge: 
\equ{
L_{g.f.} = - \frac 12 \tr (\der_M A^M)^2 = 
-\frac 12 \tr (\der_\gm A_\gm)^2  + \tr (\der_i A_\ui)^2 
+ \tr (\der_\ui A_i)^2  
\non \\ 
+ 2 \tr (\der_\gm A_\gm) (\der_\ui A_i + \der_i A_\ui) 
+ 2 \tr (\der_i A_\ui) (\der_\uj A_j).
} 
The resulting ghost action is 
\equ{
L_{ghost} = \tr \der_M\bget  D^M \get  
= \tr \der_\gm \bget D^\gm \get  + 
\tr \der_i \bget D_\ui \get  + \tr \der_\ui \bget D_i\get.
}
This completes our description of the ten dimensional gauge theory,
decomposed into an $N=1$ four dimensional language.

\subsection{Twisted fixed point states}
\labl{sc:InclTwisted}

In addition to the requirement of modular invariance, which resulted in
the stringent conditions \eqref{LevelMatching}, string theory also
gives definite predictions of the states present at the orbifold fixed
points. These twisted states can be thought of as originally open
strings, which only become closed upon non--trivial orbifold twist
identifications. For the $\mathbb{Z}_3$ orbifold this leads to the
following spectrum of chiral multiplets  
\equ{
(\rep{1}_H, \rep{S}_s: 
(w^I + v_s^I)^2 =  \mbox{\small $\frac 43$} ), 
\qquad 
(\rep{\bar 3}_H, \rep{T}_s: 
(w^I + v_s^I)^2 =  \mbox{\small $\frac 23$} ), 
\labl{twistedmatter} 
}
at fixed point $\fZ_s$. 

In the previous subsection the ten dimensional super Yang--Mills action
has been (partly) decomposed into four dimensional states. Only the
four dimensional $N=1$ supersymmetry, which is preserved by the
orbifolding, was left manifest. 
This four dimensional $N=1$ language was extended to an off--shell
formulation, involving the auxiliary fields $f_\ui$, $f_i$ and $\cD$
as functions of the ten dimensional coordinates. Therefore, the
standard rules of $N=1$ multiplet calculus 
can be used to obtain the action for the twisted chiral multiplets
$(c_s, \gps_{s\,L}, h_s)$, residing at fixed point $\fZ_s$ in the
representations \eqref{twistedmatter}, and their interactions with the
off--shell untwisted multiplets \eqref{BulkMultiplets}. For the
purpose of the tadpole calculations later, it is sufficient to give
only the scalar part of their action 
\equ{
S_{tw} = \int
\left( -D_\gm \bar c_s D^\gm c_s + \bh_s h_s - \bar c_s \cD c_s 
+ \ldots  
\right) 
\gd^6(z - \fZ_s - \gG) \, \d^6 z \, \d^4 x.
\labl{FixPointLagr}
}
The dots here may represent $F$--term contributions linear in the
auxiliary fields $f_\ui$, $h_s$ and their conjugates. We have
introduced the delta function on a fixed point of the orbifold, which
satisfies 
\equ{
\gd(z - \gth^{-k}z - \gG) = \frac 1{27} \sum_s \gd^6(z - \fZ_s - \gG), 
\qquad 
k = 1,2. 
\labl{OrbiDelta} 
}

\subsection{Models with anomalous $\boldsymbol{\U{1}}$'s}
\labl{sc:AnomModels} 

In a previous publication \cite{Gmeiner:2002es} we have computed the
local anomalies at the fixed points of the orbifold $T^6/\Intr_3$. We
found that the anomaly at fixed point $\fZ_s$ was fully determined by
the local spectrum at this fixed point (given by \eqref{LocReps} and
\eqref{twistedmatter}), and hence ultimately by its local shift
vector $v_s$. This naturally leads to the introduction of the concept
of fixed point equivalent 
models, which allows one to identify the local spectrum of this model 
at fixed point $\fZ_s$, with the spectrum of a pure orbifold model
(i.e.\ with no Wilson lines) with shift vector $v_s$. The advantage of
this is, that only a few inequivalent pure $\Intr_3$ orbifold models
exist. Therefore, the investigation of local anomalies reduces to
the analysis of those pure orbifold models. 

The full four dimensional anomaly $I^1_{4|s}$ at fixed point
$\fZ_s$, is given by the solution \eqref{GenAnomaly} (c.f. appendix
\ref{sc:AnomPolyFact}) of the descent equation from the anomaly
polynomial 
\equ{
I_{6|s} = 
\Bigl\{
\frac 3{27} \ch{\rep{R}_s}{iF_2}
+ 3 \ch{\rep{T}_s}{iF_2}
+ \ch{\rep{S}_s}{iF_2}
\Bigr\}
\hat {\text{A}}[R_2] \Big|_{6|s},
\labl{AnomPoly6pre}
}
where wedge products are implicitly understood. The subscript $6|s$
indicates that this formal expression is restricted to the 6--form
part, and refers to the anomaly at fixed point 
$\fZ_s$. This requires that both the field strength $F_2$ of
$\E{8}\times\E{8}'$ and the curvature 2--form $R_2$
are restricted to this fixed point. Here the Chern character
$\text{ch}_{\rep{r}}[iF_2] = \tr_{\rep{r}} \exp (i F_2/2\pi)$ is
computed in representation $\rep{r}$, and $\hat{\text{A}}(R_2)$
denotes the roof (Dirac) genus. (For an exposition of some useful
properties of (Chern) characters, see appendix \ref{sc:char}.) 
We have used that pure gravitational anomalies can never arise in
four dimensions. The factor of $1/27$ appears because the bulk fields  
constitute at a given fixed point of $T^6/\Intr_3$ only $1/27$ part of
the usual anomaly.  

As was shown in \cite{Gmeiner:2002es} using fixed point equivalent
models, the non--Abelian gauge anomalies always cancel. Therefore we
only need to  consider the possible Abelian anomalies (both pure and
mixed) more carefully. At a given fixed point $\fZ_s$ we may have at
most one anomalous $\U{1}$. Its gauge field is denoted by $A_1'|_s$,
while the other gauge fields that exist at this fixed point are
collectively referred to as $\tA_1|_s$. Employing similar notation for
the corresponding field strengths, the anomaly polynomial becomes  
\equ{
I_{6|s} = 
\frac i{48} \tr_{\rep{L}_s} \Bigl( \frac{F_2'}{2\gp}\Bigr)  
\Bigl.\tr \Bigl( \frac {R_2}{2\gp} \Bigr)^2\Bigr|_s -  i\,
\tr_{\rep{L}_s} \Bigl[
\frac 16 \Bigl( \frac{F_2'}{2\gp} \Bigr)^3 + 
\frac 12 \frac{F_2'}{2\gp}\, \Bigl( \frac{\tF_2}{2\gp} \Bigr)^2
\Bigr]_s
\labl{AnomPoly6}
}
Here we have utilized the symbolic short--hand notation 
$\rep{L}_s = \frac 3{27} \rep{R}_s + \rep{S}_s + 3 \rep{T}_s$ to
denote the local matter representations with their relevant
normalizations at fixed point $\fZ_s$. In addition, from the fixed
point model analysis it followed, that there are 
only two types of anomalous $\U{1}$'s at a given fixed point
\cite{Gmeiner:2002es}. The relevant $\E{8}$ branching rules
\cite{Slansky:1981yr} are given by 
\equ{
\arry{l}{
\arry{lclcl}{
\E{8} &\ra& \E{7} \times \SU{2} &\ra& \E{7} \times \U{1}, 
\\[0ex]
\rep{248} &\ra& (\rep{133}, \rep{1}) + (\rep{1}, \rep{3}) 
+ (\rep{56}, \rep{2}) 
&\ra& \rep{133}_0 + \rep{1}_0 
+  \rep{1}_{2} + \rep{1}_{\mbox{-}2} 
+ \rep{56}_{1} + \rep{56}_{\mbox{-}1} ;  
}
\\[2ex]
\arry{lclcl}{
\E{8} &\ra& \SO{16}  &\ra& \SO{14} \times \U{1}' , 
\\[0ex]
\rep{248} &\ra& \rep{120} +  \rep{128} 
&\ra& \rep{91}_0 + \rep{1}_0 
+  \crep{14}_{2} + \rep{14}_{\mbox{-}2} 
+ \rep{64}_{1} + \crep{64}_{\mbox{-}1} ;
}
\\[2ex]
\arry{lcl}{
\E{8} &\ra& \SU{9} , 
\\[0ex]
\rep{248} &\ra& \rep{80} + \rep{84} +  \crep{84}. 
}
}
\labl{branchings}
}
The spectra corresponding to the pure orbifold models have been
summarized in table \ref{tab:AnomFixedModels}. The last column of this
table gives the traces over $\rep{L}_s$ of the possible 
anomalous charges $(q_s, q_s')$. Since $q_s$ in the $\SU{9}$ model is
part of the generators of $\SU{9}$ it is of course tracesless over
each representation. The one but last column gives the traces over the
untwisted, bulk, representations $\rep{R}_s$. As can be deduced by
comparing these two final columns, in the $\E{7}$ model the generator
$q_s'$ of the $\U{1}'$ is traceless, and therefore anomaly free, only
if both untwisted and twisted states are taken into account. 
The anomalous $\U{1}$'s of these models are  ``universal'', in the
sense that the following relation holds \cite{Schellekens:1987xh}  
\equ{
\frac 16\, \frac{1}{k_{q_s}}\tr_{\rep{L}_s} 
\bigr( q_s^3 \bigr) = 
\frac 14 \sum_{a} q_s({\rep{L}_s^{(a)}}) \, I_2(\rep{L}_s^{(a)})
= 
\frac{1}{48}\, \tr_{\rep{L_s}} \bigl( q_s \bigr). 
\labl{U1universality}
}
The sum is over the irreducible representations $\rep{L}_s^{(a)}$
of the gauge group factors $\smash{G_{(a)}}$ in $G_s$.  
The quadratic indices $\smash{I_2(\rep{L}_s^{(a)})}$ are normalized
w.r.t.\ to these factor groups, and $\smash{q_s({\rep{L}_s^{(a)}})}$ 
is the $\U{1}$ charge of $\smash{\rep{L}_s^{(a)}}$. (For a more detailed 
discussion of the indices and their normalization, see appendix
\ref{sc:E8traces}, and refs.\
\cite{Schellekens:1987xh,Kobayashi:1997pb,vanRitbergen:1998pn}.) 
Because of the inclusion of the level  
\(
k_{q_s} = 2q_{s}^2
\)
of $q_s$ this formula is valid for any normalization of this local
$\U{1}$ generator.

\begin{table}
\rotatebox{0}{\scalebox{.85}{
\renewcommand{\arraystretch}{1.25}
  \begin{tabular}{|l|l|l|ll|c|c|}\hline
    {Model }
      & {Shift $v_s$ and}
      & {Untwisted}
      & {Twisted}
      & 
& { $\frac 3{27}\tr_{\rep{R}_s}$ of }
& {~~$\tr_{\rep{L}_s}$ of~~}
\\
& {gauge group $G_s$} 
& {$(\rep{3}_H,\rep{R}_s)$ }
& {$(\rep{1}_H, \rep{S}_s)$ } 
& {$(\rep{3}_H, \rep{T}_s)$}
& {$(q_s, q_s')$}
& {$(q_s, q_s')$}
      \\\hline
    {$\E{7}$}
      & {$\frac{1}{3}\!\left(~0,~1^2,~0^5 ~|~ \mbox{-}2,~0^7 \right)$}
      & {$(\rep{1})_{0}(\rep{64})'_{\frac{1}{2}}
 + (\rep{56})_{1}(\rep{1})'_{0}$}
      & {$(\rep{1})_{\frac{2}{3}}(\rep{14})'_{\mbox{-}\frac{1}{3}} $}
      & {$ (\rep{1})_{\frac{2}{3}}(\rep{1})'_{\frac{2}{3}}$}
& {$(6, 2)$}
&  {$(16, 0)$} 
      \\
      &  $  { \E{7}\!\times\!\U{1} \times
\SO{14}'\! \times\! \U{1}'} \!\!$
      & {$+\, (\rep{1})_{0}(\rep{14})'_{\mbox{-}1} + (\rep{1})_{\mbox{-}2}(\rep{1})'_{0}$}
      & {$ +\,  (\rep{1})_{\mbox{-}\frac{4}{3}}(\rep{1})'_{\frac{2}{3}}$ }
      & & &
      \\\hline
    {$\SU{9}$}
     & {$ \frac{1}{3}\! \left(\mbox{-}2,~1^4~,0^3 ~|~ \mbox{-}2,~0^7 \right) $}
     & {$(\rep{84})(\rep{1})'_{0} + (\rep{1})(\rep{64})'_{\frac{1}{2}}$}
      & {$(\crep{9})(\rep{1})'_{\frac{2}{3}}$}
     & 
& {$(0,2)$}
& {$(0, 8)$}  
      \\
      & ${\SU{9} \times \SO{14}'\!\times\!\U{1}'} \!\!$
      &{$+\, (\rep{1})(\rep{14})'_{\mbox{-}1}$} & & & &
      \\\hline
    \end{tabular}
}}
\caption{This table gives the gauge groups $G_s$, the untwisted
($\rep{R}_s$) and twisted matter ($\rep{S}_s$ and $\rep{T}_s$)
representations of the pure orbifold models with an anomalous $\U{1}$
factor. The $\E{7}$ model contains two $\U{1}$ factors, of which only the first
one is anomalous. With the sign conventions of the charges as in this
table both traces for the anomalous generators are positive.}
\labl{tab:AnomFixedModels}
\end{table}

\section{Green--Schwarz mechanism on the orbifold 
$\boldsymbol{T^6/\Intr_3}$}
\labl{sec:GreenSchwarz} 

In \cite{Gmeiner:2002es} we have derived the full structure 
of the gauge anomaly on $T^6/\Intr_3$. In a similar fashion, also pure  
gravitational and mixed gauge--gravitational anomalies can be obtained.
The full ten dimensional anomaly of this orbifolded theory is given by 
\equ{
\int A_{10}^1 = \int  \frac 13 \, I_{10}^1 
+ \sum_{s} I^1_{4|s} \, \gd^6(z-\fZ_s - \gG) \, \d^6 z. 
\labl{AnomStruc}
} 
The factor $1/3$ results from the orbifold projection; only $1/3$ of
the ten dimensional states on the torus $T^6$ survive the orbifold twist.
The anomaly $I_{10}^1$ is determined by the descent equations
\eqref{Descent} of appendix \ref{sc:AnomPolyFact}  
from the anomaly polynomial 
\equ{
I_{12} = \hat{\text{A}}_{3/2}[R_2] - \hat{\text{A}}[R_2] +
\text{ch}_{\E{8}\!\times\!\E{8}}[iF_2]\,  
\hat {\text{A}}[R_2] \,  \Big|_{12}.
\labl{AnomPoly12}
}
The first term results from the left--handed 
(spin $3/2$) gravitino $\gps_M$, the second term is due to the
right--handed dilatino $\gl$, and the final term is the consequence of the
gaugino $\gch$ of the $\E{8}\times \E{8}'$ super Yang--Mills gauge
theory. The anomaly polynomial $I_{6|s}$ of the four dimensional
anomaly $I_{4|s}^1$ at fixed point $\fZ_s$ has already been discussed
in section \ref{sc:AnomModels}.  As the non--Abelian anomalies cancel,
$I_{6|s}$ reduces to \eqref{AnomPoly6} and is non--vanishing, only if the
spectrum at this fixed point is equivalent to the $\E{7}$ or
$\SU{9}$ spectra (given in table \ref{tab:AnomFixedModels}). 
The aim of this section is to show, that the ten dimensional anomaly
and the four
dimensional Abelian anomalies at the fixed points can be canceled
simultaneously by an anomalous variation of the anti--symmetric tensor.

The theory of $N=1$ supergravity in ten dimensions has two equivalent
formulations, using either the anti--symmetric tensors residing in the 
2--form $B_2$, or the 6--form $C_6$ potential  
\cite{Chamseddine:1981ez,Bergshoeff:1982um,Chapline:1983ww}. 
Their 1--form and 5--form gauge transformations  
$\gd_{\gL_1} B_2 = \d \gL_1$ and $\gd_{\gL_5} C_6 = \d \gL_5$ leave 
their actions 
\equ{
\arry{l}{ \dsp  
S_2 = \int   - \frac 12 *\d B_2 \, \d B_2 + (*X_3 + X_7 ) \d B_2  
- \frac 12 * X_3 \, X_3, 
\\[2ex] \dsp 
S_6 = \int  
- \frac{1}{2} *(\d C_6 + *X_3 + X_7)  (\d C_6 + * X_3 +  X_7)
 - \frac 12 * X_3 \, X_3,
}
\labl{LagrBC}
}
invariant. Here the 3-- and 7--forms $X_3, X_7$ are derived from 
arbitrary closed 4-- and 8--forms, $X_4, X_8$, by Poincar\'e's lemma  
(i.e.\ we have locally $\d X_3 = X_4$ and $\d X_7 = X_8$). 
To show that these two actions are equivalent, start with $S_2$
for example: Introduce the canonical field strength 
$H_3 = \d B_2 - X_3 - *X_7$ and a 6--form Lagrange multiplier $C_6$ to
enforce the Bianchi identity $\d H_3 + X_4 + \d * X_7 = 0$. Eliminate
field strength $H_3$, using its algebraic equation of motion, to obtain
action $S_6$.  

We now determine for which $X_3$ and $X_7$ the variations of these
actions under gauge and local Lorentz transformations, with
infinitesimal parameters $\gL$ and $L$, respectively, cancel all 
anomalies of \eqref{AnomStruc}. The non--Abelian variations of the
gauge connection $A_1$ and the spin--connection $\go_1$
\equ{
\gd_\gL A_1 = \d \gL + [\gL, A_1],
\qquad 
\gd_L \go_1 = \d L + [L, \go_1],
}
lead to the transformations  
\equ{
\gd_\gL B_2 = X_2^1, 
\qquad 
\gd_\gL C_6 = - X_6^1, 
}
and similarly for $L$. 
Here we have assumed that $X_4$ and $X_8$ are gauge invariant; 
hence $\gd X_3 = \d X_2^1$ and $\gd X_7 = \d X_6^1$ locally. 
These variations follow because the anomaly \eqref{AnomStruc}, and
hence the variations of the actions \eqref{LagrBC}, do not contain any
dependence on either of these higher tensor fields. Since the anomaly
\eqref{AnomStruc} does not contain any Hodge dualization, the final
term  $ - \frac 12 * X_3 X_3$ is needed in \eqref{LagrBC}. 
Therefore, the variations of both actions  $S_2$ and $S_6$ are equal
to 
\equ{
\gd_\gL S_2 = \gd_\gL S_6 = \int X_7\, \gd_\gL X_3.
\labl{VarActASTens}
}
Since the ten dimensional part of the anomaly on the orbifold
\eqref{AnomStruc} is one third of the anomaly in ten dimensional
Minkowski space, we expect that the original Green--Schwarz mechanism
will be relevant here as well. For that reason we briefly review it
here. 

The crucial observation by Green and Schwarz \cite{Green:1984sg} for 
anomaly cancellation in ten uncompact dimensions is, that
the anomaly can be cancelled if the corresponding anomaly polynomial 
\eqref{AnomPoly12} factorizes: 
\equ{
I_{12} = X_{4\, GS} X_{8\, GS}, 
\labl{GrSchw12}
}
see \cite{gsw_2}. The normalization of $X_{4\, GS}$ is fixed by 
supersymmetry\footnote{
This has only been explicitly checked for the Yang--Mills part 
because the Lorentz part is of higher order in derivatives.
}
since the
gauge Chern--Simons term $\go_{3\, Y}$ appears in the supergravity
Lagrangian when it is coupled to super Yang--Mills theory, hence 
\equ{
X_{3\, GS} = \go_{3\, L} - \frac 1{30} \go_{3\, Y}, 
\qquad 
X_{4\, GS} = \d X_{3\, GS}=  \tr \, R^2_2 - \frac 1{30} \, \Tr F_2^2,
}
with the standard notation $\Tr = \tr_{\E{8}\!\times\!\E{8}'}$ for the 
trace in the adjoint of $\E{8}\!\times\!\E{8}'$. 
(The factor of $1/30$ can be thought of as the normalization of the
$\tr_{\E{8}\!\times\!\E{8}'}$ trace in $\go_{3\, Y}$ w.r.t.\ the
vector representation of $SO(1,9)$.)  For the group 
$\E{8}\times\E{8}'$  the factorization equation 
\eqref{GrSchw12} can be satisfied, with 
\equ{
X_{8\, GS} = \frac 1{(2\gp)^5}\left[
\frac 1{24} \Tr F_2^4 - \frac 1{7200} (\Tr F_2^2)^2 
- \frac 1{240} \Tr F_2^2 \tr R^2 + \frac 18 \tr R^4 
+ \frac 1{32} (\tr R^2)^2
\right].
}
With these ingredients we return to the local anomaly cancellation
on the orbifold.

As the $T^6/\Intr_3$ anomaly is $\frac 13$ of the original ten dimensional
anomaly, and the Chern--Simons term $\go_{3\, Y}$ in the field strength
$H_3$ is required by the ten dimensional supersymmetry in the bulk, 
we infer that $X_7 = \frac 13\,  X_{7\, GS} + \ldots$ and 
$X_3 = X_{3\, GS} + \ldots$; the dots refer to additional terms 
which are relevant for the cancellation of the four dimensional fixed
point anomalies. Since the fixed points of the orbifold $T^6/\Intr_3$
are isolated and have codimension six, the corresponding orbifold
delta function cannot be factorized, and hence should reside completely
within $X_7$. (Otherwise, we would have lower dimensional hyper
planes.)  Hence, we conclude that 
\equ{
X_3 = X_{3\, GS}, 
\qquad 
X_7 = \ga\,  X_{7\, GS} + \sum_s \ga_s \, A_1' |_s 
\gd^6(z-\fZ_s - \gG) \d^6 z,
\labl{3and7forms}
}
with $\ga = \frac 13$ and $\ga_s$ some constants. 
As the fixed point anomalies only involve mixed and pure $\U{1}$
anomalies of the anomalous $\U{1}$'s, the 
anomaly polynomials $I_{6|s}$ have to factorize like 
\equ{
I_{6|s} = \ga_s \, X_{4\, GS}|^{}_{s} F_2'|^{}_s, 
\qquad 
\left. X_{4\, GS}\right|_s = 
\tr R_2^2|_{s}^{} - 2 \sum_a \tr F^2_{(a)}|_{s}^{}\,.
\labl{Fact6}
}
Here  $X_{4\, GS}|_s$ denote the restrictions of $X_{4\, GS}$ to the
groups $G_s$ present at the fixed point $\fZ_s$, the sum is over the
gauge  
group factors in $G_s$, and traces $\tr F^2_{(a)}$ are normalized 
with respect to the quadratic indices of the respective gauge group
factors. For details and a proof of the second equation in
\eqref{Fact6}, we refer the reader to appendix \ref{sc:E8traces}, 
where the relevant calculations are performed. Now, precisely
because of the  ``universality'' relation \eqref{U1universality} for
the anomalous $\U{1}$'s of the two anomalous pure orbifold models, the
above expression for $I_{6|s}$ is proportional to that given by eq.\ 
\eqref{AnomPoly6}. The normalization factor is easily determined to be 
$\ga_s = \frac 1{48} \frac 1{(2\gp)^3}$ and turns out to be fixed
point independent. The full Green--Schwarz
action reads  
\equ{
S_{GS} =  S_*  + 
\Bigl( 
\gb\, X_{7\, GS} + 
 \sum_s \gb_s\, A_1' |_s  \, \gd^6(z-\fZ_s) \d^6 z 
\Bigr) X_{3\, GS},
}
where $S_* $ may either be $S_2$ or $S_6$, depending whether one uses
the  2-- or 6--form formulation of supergravity. The coefficients
$\gb$ and $\gb_s$ are determined below. The gauge variation
of this action is given by  
\equ{
\arry{rcl}{
\gd_\gL S_{GS} &=&  \dsp \int \gb\, \d X^1_{6\, GS} \, X_{3\, GS} 
+ ( \ga + \gb) X_{7\, GS} \, \d X^1_{2\, GS}  
\\[2ex]
& & \dsp  \qquad+ 
\Bigl( 
 \gb_s\, \gL^s  X_{3\, GS} 
+ 
(\ga_s + \gb_s)\,  A_1'|_s
\d X^1_{2\, GS}.
\Bigr) 
\,  \gd^6(z-\fZ_s) \d^6 z. 
} 
}
Here we have used equation \eqref{FactAnom} of appendix
\ref{sc:AnomPolyFact} to determine  the actual form of
the anomaly given by the factorization of the anomaly polynomials
\cite{Green:1985bx}. 
This, in fact, fixes the coefficients: $\gb = - 2/3$ and 
$\gb_s = \ga_s$.

We close this section with some comments to link these results to the
well--known situation of the zero mode theory of heterotic models 
on $T^6/\Intr_3$ with an anomalous $\U{1}$. As
discussed in \cite{Schellekens:1987xh,Kobayashi:1997pb} for all 
orbifold models with Wilson lines the ``universality'' relation 
\eqref{U1universality} holds if the model contains an anomalous
$\U{1}$.  With the local anomaly cancellation
presented here, this result can be understood easily: For the two
pure orbifold models with an anomalous $\U{1}$ (the $\E{7}$ and
$\SU{9}$ models) this relation holds; hence it holds for all localized
anomalous $\U{1}$'s in all $\Intr_3$ models with Wilson lines as
well, since the model at an `anomalous' fixed point is equivalent to one of the
two pure orbifold models with an anomalous $\U{1}$.
Moreover, we know that the zero mode anomalous $\U{1}$ is a
linear combination of the local anomalous $\U{1}$'s, see 
\cite{Gmeiner:2002es}. Therefore, also the anomaly of the zero mode
$\U{1}$ is canceled, and zero mode factorization is implied. 

Finally, we turn to the issue of the model independent axion(s). 
Notice that the second term of the second equation in \eqref{3and7forms} 
leads to the interaction (in the 2--form formulation)
\equ{
\int \ga_s \, A_1'|_s\, \gd^6(z-\fZ_s - \gG)\d^6 z \, \d B = 
\sum_s \int \ga_s \, A_\gm' |_s \der_\gm b_s \, 
\gd^6(z- \fZ_s - \gG)\d^6 z\, \d^4 x,
}
of the local anomalous $\U{1}$ gauge field $A_\gm'|_s$ and the
anti--symmetric tensor. (This coupling is precisely the
local version of the zero mode interaction $A_\gm \der^\gm b$,
discussed in  \cite{Dine:1987xk}.)  Because of the delta function
$\gd^6(z-\fZ_s)\d^6 z$ the exterior derivative on $B$ acts only 
in the four non--compact dimensions: $\d = \d_{(4)}$. (The subscript
$_{(4)}$ emphasizes, that manipulations like Hodge dualization and
exterior differentiation are performed in four dimensions.) 
By performing local dualization in four dimensions 
we have introduced the fixed point axions $b_s$ by 
\equ{
\d_{(4)} B_2(x, z)|_s = \d_{(4)} B_2(x, \fZ_s) = *_{(4)} \d_{(4)} b_s(x).  
}
Notice, that these fixed point axions $b_s$ are only defined on the fixed
points. The model independent axion $b(x)$ is the dual of the
zero mode of the four dimensional anti--symmetric tensor
$B_{\gm\gn}(x)$. After substituting this in the above equation and
considering the zero mode anomalous $\U{1}$ gauge field, it follows
that the model independent axion is identified as the sum of
all local axions: $b = \sum_s b_s$.

\section{Tadpoles}
\labl{sc:Tadpoles}

In four dimensional $N=1$ supersymmetric $\U{1}$ gauge
theories coupled to chiral multiplets, one can show that the auxiliary
field of the gauge multiplet acquires a quadratically divergent
tadpole at one loop, which is proportional to the sum of charges 
\cite{Fischler:1981zk}. In section \ref{sc:EffSYM} we showed that also
the full ten dimensional super Yang--Mills theory can be cast in the
form of an $N=1$ off--shell theory in four dimensions. Therefore, it
is natural to consider the possibility of tadpoles for the auxiliary
fields $\cD$ introduced in section \ref{sc:4Dsusy}. These tadpoles are 
computed in the next subsection. Because of the supersymmetry structure
of the $D$--term scalar potential in \eqref{6DYM}, one would expect
that a tadpole for $\cD$ arises if and only if there is also a tadpole 
for $\der_i A_\ui$. Therefore (as a cross check) we compute 
all tadpoles for $A_\ui$ in section \ref{sec:TadpolesIntGauge}.

Before, we turn to the details of these tadpole calculations, we first
describe the basic technique to perform the loop calculation for the
bulk states on the orbifold $T^6/\Intr_3$. Obviously, it is much
more convenient to perform the whole computation on the torus $T^6$.
To take the orbifolding into account we insert an explicit orbifold
projection operator that projects onto orbifold covariant
states. (This method has been applied for the related anomaly
calculation in ref.\ \cite{Gmeiner:2002es}, see also ref.\
\cite{Poppitz:1998dj} for a string computation of Fayet--Iliopoulos
tadpoles in type I models.) To explain the method, consider an operator
$\cO(z)$ that acts on a Hilbert space associated to a scalar field
$S$ on the torus $T^6$. Let $\{\gf_q(z)\}$ be an orthonormal basis for
this Hilbert space. For example, the basis \eqref{TorusWave} defined in
appendix \ref{sc:WaveProp}, can be used. However, it should be
stressed that our results are independent of the basis chosen for this
Hilbert space. The expectation value of $\cO(z)$ on the
torus reads  
\equ{
\langle \cO(z) \rangle_{T^6} = 
\sum_q \gf_q^\dag(z) \, \cO(z) \, \gf_q(z). 
}
For the computation on the orbifold, it is essential that this scalar
$S$ transforms covariantly under the orbifold twist: 
\(
S(\gTh z) = \gth^\gs \, S(z), 
\)
where the eigenvalue $\gs = 0, 1, 2$ is defined modulo $3$. The
expectation value of the same operator on the orbifold is defined as 
\equ{
\langle \cO(z) \rangle_{T^6/\Intr_3} = 
\frac 1{3} \sum_k \gth^{-\gs\,  k}
\gf_{q }^\dag (\gth^{-k} z) \, \cO(z)\, \gf_{q}(z). 
\labl{OrbiProjc}
}
The part of this expression with $k=0$ gives the same
contribution as the torus expectation value, up to the 
normalization factor $\frac 13$. It is straightforward to extend this
procedure to other fields on the orbifold in more complicated
representations. In the tadpole calculations below we apply this
technique to the homogeneous twist components of the gauge fields,
gauginos and ghosts. Their symmetrization factors have been collected
in table \ref{tab:SymmFact}. 

Before we turn to the explicit tadpole calculation, we make one
technical comment: All our integrals are taken over Euclidean momentum
space; i.e.\ the Wick rotation from Minkowskian momentum space has
been performed implicitly.

\begin{table}
\renewcommand{\arraystretch}{1.25}
\begin{center} 
\tabu{|l|l|l|}{
\hline 
State & $\gs(\ldots)$ & ~~ Value  
\\\hline 
Gauge field & $\gs(M, w)$ &
\(
\begin{cases} 
3 v^I w_I & M = \gm  \\
2 + 3 v^I w_I & M = i   \\
1 + 3 v^I w_I & M = \ubar i   
\end{cases}
\)
\\\hline 
Gaugino & $\gs(\gk, w)$ & $~~~ 3( \frac 12 \gf^i \gk_i + v^I w_I) $
\\\hline 
Ghost & $\gs$ & $~~~3 v^I w_I$
\\\hline 
}
\end{center} 
\caption{This table gives the
symmetrization factors $\gs(\ldots)$ needed for the computation of the
tadpole diagrams in dependence of the fields and their components.}
\labl{tab:SymmFact}
\end{table}

\subsection{Fayet--Iliopoulos tadpole for $\boldsymbol{\cD}$}
\labl{sc:FItad}

In figure \ref{fig:FItadp} we have given the possible tadpole diagrams
for $\cD$.  We have employed the following notation for the 
 first diagram of figure \ref{fig:FItadp}, which has internal gauge
fields $A_\uj$ in the loop (corresponding to the second term in the
second line of equation \eqref{6DYM}). The dotted lines refer to gauge
index contractions using the inverse Killing metric $\get_{\ga\gb}$ 
and a vertex of dotted lines refers to the structure 
coefficient $f_{\ga\gb\gg}$. This means that loops of dotted lines
imply that we have to sum over all generators of $\E{8}\times \E{8}'$. 
The solid lines refer to contractions of spacetime indices.  
Since $T^6$ is described as a complex manifold, these solid lines
carry a complex orientation, which we indicate using open arrows. In
the second diagram of figure \ref{fig:FItadp},  
the fixed point twisted scalars $c_s$ run around in the loop, see 
the interaction term in \eqref{FixPointLagr}.

\begin{figure}
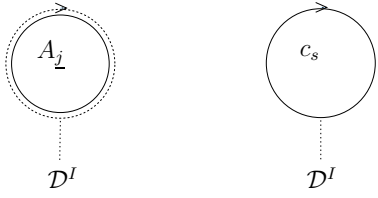

\[
\arry{ccc}{
\raisebox{0ex}{\scalebox{0.8}{\mbox{\input{FI_bulk.pstex_t}}}}
%\qquad & \qquad 
%\raisebox{0ex}{\scalebox{0.8}{\mbox{\input{FI_ghost.pstex_t}}}}
\qquad & \qquad 
\raisebox{0ex}{\scalebox{0.8}{\mbox{\input{FI_brane.pstex_t}}}}
}
\]
\caption{The following diagrams give rise to FI--tadpoles of $\cD$. 
In the loop we have internal gauge fields $A_\uj$, and fixed point
states $c_s$.} 
\labl{fig:FItadp}
\end{figure}

Only auxiliary $\cD$ with a Cartan gauge index
($I$) can develop a tadpole: The propagators are diagonal in the gauge
indices, therefore, it is not possible
to form a closed tadpole diagram with a root index ($w$) on the
external line. Since both the Wilson lines and the orbifold twist 
are generated by the Cartan subalgebra, such tadpoles are allowed by
the boundary conditions of the heterotic orbifold theory.

The diagram with the internal gauge fields (untwisted states) in the
loop gives rise to  
\equ{
\gx_{I\,un} =  
\frac 1{3} \int \frac{\d^4 p}{(2\gp)^4} 
~\, \sum_{k, w, q,\uj}~\, \gth^{-\gs(\uj, w)k} ~~
\gf_{qw}^\dag(\gth^{-k} z) \, \frac {1}{p^2 + \gD}\,  \gf_{qw}(z) 
\, f_{Iw}^{~~~w}, 
}
where we have introduced the internal Laplacian 
$\gD = - 2 \sum \der_i \der_\ui$. The case $k = 0$ does not
contribute, since it is proportional to the trace of the Cartan
generator $H_I$ over the full adjoint of $\E{8}\times \E{8}'$. 
For $k \neq 0$ we would like to rewrite the sum over mode functions 
as fixed point delta functions, using identity \eqref{CombiWaves} 
of appendix \ref{sc:WaveProp}. Clearly, we are only able to do so, 
if the Laplacian acts on the product of the mode functions $\gf_{qw}$. 
This can be achieved with the help of \eqref{DerProducts} of the same
appendix, and we obtain 
\equ{
\gx_{I\,un} =  
\frac 1{3} 
~\, \sum_{k, s,w, \uj} 
\,  ~~
\gth^{-\gs(\uj,w,s)k} ~ f_{Iw}^{~~~w} ~
 \, \int \frac{\d^4 p}{(2\gp)^4} \frac {1}{p^2 + \frac 13 \gD}\, 
\frac 1{27} \gd(z - \fZ_s - \gG). 
\labl{subresultFI}
}
Here we have introduced the fixed point dependent symmetrization
factor $\gs(\uj, w, s) = 1 + 3 v_s^I w_I$ corresponding to the local
shift vector $v_s$ at the fixed point $\fZ_s$. This can be rewritten
further, as a sum over the local representations 
$\rep{r} = \rep{R}_s, \crep{R}_s$ and $\rep{Ad}_s$ defined in
\eqref{LocReps}.  
Of course, the trace of $H_I$ in the local adjoint $\rep{Ad}_s$
vanishes, and hence will be dropped. Furthermore, we have
that   
\(
\tr_{\rep{R}_s} (H_I) = -\tr_{\crep{R}_s} (H_I) 
= \sum_{w\in \rep{R}_s} f_{Iw}^{~~~w}.
\) 
We use the notation: $(-)^\rep{r} = +,-$ for 
$\rep{r} = \rep{R}_s$ and $\rep{r} =  \crep{R}_s$, respectively. 
With these definitions the expression above becomes  
\equ{
\gx_{I\,un} =  
\frac 1{3} 
\sum_{k, s,\uj} 
\tr_{\rep{R}_s}(H_I)
\sum_{\rep{r} = \rep{R}_s, \crep{R}_s } 
\gth^{-\gs(\uj, \rep{r})k} 
\, 
(-)^{\rep{r}}
\, 
 \int \frac{\d^4 p}{(2\gp)^4} \frac {1}{p^2 + \frac 13 \gD}\, 
\frac 1{27} \gd(z - \fZ_s - \gG). 
\labl{subresultFI2}
}
To evaluate this we need to compute the sum: 
\equ{
\sum_{k=1,2} \sum_{\rep{r} = \rep{R}_s, \crep{R}_s}
\gth^{- \gs(\uj, \rep{r})k} (-)^{\rep{r}}
= 3(2 -  \gth^2  - \gth) = 9.
\labl{SymmSurf}
}
By Taylor expanding to first order in  $\frac 13 \gD$, and performing
the resulting divergent integrals using the cut--off scheme, we find 
that the FI--parameter takes the form  
\equ{
\gx_{I\,un} =  \sum_s
\, 3 \tr_{\rep{R}_s}(H_I) \, 
 \left( 
\frac {\gL^2}{16 \gp^2} 
+ \frac {\ln \gL^2}{16 \gp^2}\,  
\frac 13 \gD 
\right) 
\frac 1{27}  \gd(z - \fZ_s - \gG),
}
where $\gL$ denotes the cut--off scale.

The brane contributions are easier to obtain, as they are already
confined to the four dimensional orbifold planes. Their effect on the 
FI--parameter can be 
read of straightforwardly from \eqref{FixPointLagr}. As the complex
scalars $c_s$ of fixed point $\fZ_s$ reside in the representations 
\eqref{twistedmatter}, their tadpole contribution reads 
\equ{
\gx_{I\, tw} = \sum_s ( \tr_{\rep{S}_s+ 3 \rep{T}_s})(H_I) \, 
\frac {\gL^2}{16 \gp^2} \,  \gd(z - \fZ_s - \gG). 
}
Combining these results, we find the expression for the 
full FI-term 
\equ{
L_{FI}=-\gx_I\cD^I\,,\qquad
\gx_{I} = \sum_s
\left( 
\frac {\gL^2}{16 \gp^2}\,  \tr_{\rep{L}_s}(H_I)
+ 
\frac 1{27} 
\frac {\ln \gL^2}{16 \gp^2}\,  \tr_{\rep{R}_s}(H_I)
 \gD 
\right) 
\gd(z - \fZ_s - \gG). 
\labl{FItadp}
}
The sign in this Fayet--Iliopoulos action is dictated by the Wick
rotation. 
Here we have again used the notation $\tr_{\rep{L}_s}$ which has been
introduced in eq.\ \eqref{AnomPoly6}. The quadratically divergent part
of the FI--parameter $\gx_I$ is proportional to precisely the same
trace which determines the localized anomalous $\U{1}$'s, see
\eqref{AnomPoly6}. The logarithmically divergent part of this
expression is proportional to the trace over the bulk states only. As
can be seen from the one but last column of table
\ref{tab:AnomFixedModels}, for all local $\U{1}$ factors, not just the
anomalous ones, this logarithmically divergent part is present. 

It is not difficult to see that this expression reduces to the
well--known result at the zero mode level, by integrating out the
internal dimensions of the orbifold. In particular the second term,
with the Laplacian $\gD$, then drops out. In fact, since we are
considering the low energy regime of a heterotic string model, the
cut--off $\gL$ should be related to the string scale. The
calculation of the zero mode Fayet--Iliopoulos terms has been performed
in full heterotic string theory, see refs.\
\cite{Atick:1987gy,Dine:1987gj}. From these calculations we infer that
$\gL = 1/\sqrt{3 \ga'}$ with $\ga'$ the string tension.

\subsection{Tadpoles of the internal gauge fields}
\labl{sec:TadpolesIntGauge}

This section is devoted to the computation of tadpoles of internal
gauge fields. As for the auxiliary fields, it is not possible to have
tadpole diagrams of internal gauge fields with non--Cartan indices. The
computation of tadpoles for $A_i^I$ and $A_\ui^I$ are completely
analogous and hence we focus on the tadpoles of $A_\ui^I$ only. As the
contributions of the internal gauge fields $A_\uj$ to the tadpoles 
of $A_\ui^I$ are rather subtle, we discuss them first and in more detail.

\begin{figure}
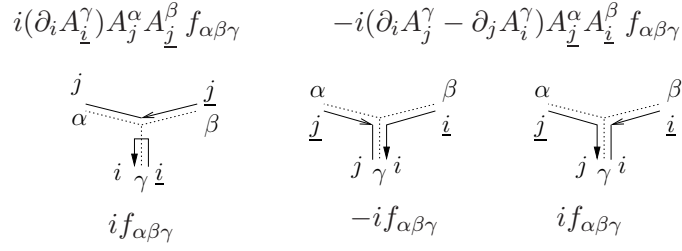

\[
\renewcommand{\arraystretch}{1.5}
\arry{cc}{
i(\der_i A_\ui^\gg) A_j^\ga A_\uj^\gb \, f_{\ga\gb\gg}
\quad 
& 
\quad 
-i(\der_i A_j^\gg - \der_j A_i^\gg) A_\uj^\ga A_\ui^\gb \, f_{\ga\gb\gg}
\\[2ex]
\arry{c}{
\raisebox{0ex}{\scalebox{0.8}{\mbox{\input{verS1.pstex_t}}}}
\\[-2mm] 
i f_{\ga\gb\gg}
}
&
\arry{cc}{
\raisebox{0ex}{\scalebox{0.8}{\mbox{\input{verT1.pstex_t}}}}
\quad & \quad  
\raisebox{0ex}{\scalebox{0.8}{\mbox{\input{verT2.pstex_t}}}}
\\[-2mm]
- i f_{\ga\gb\gg} & i f_{\ga\gb\gg} 
}
}
\]
\caption{The vertices relevant for the tadpoles of
$A_\ui^I$ involving internal gauge fields $A_\uj^w$ and $A_j^w$. The
adjoint indices $\ga, \gb$ and $\gg$ can refer to both Cartan indices
$I$ as well as root $w$.}
\labl{fig:Vertices}
\end{figure}

Only three cubic terms in \eqref{6DYMpre} are relevant for scalar
contributions to the tadpole of $A_\ui^I$. (The reason is, that for 
cubic terms with two $A_j$ one cannot close the loop if $A_\ui$
represents an external leg.) In figure \ref{fig:Vertices} we
have collected these three terms and drawn the corresponding
vertices.  In addition to the Feynman rules introduced in the previous
section, a solid arrow at the end of a solid line indicates
differentiation w.r.t.\ a holomorphic coordinate $z_i$.  
Using these vertices one can draw four different diagrams which give
rise to tadpole contributions of $A_\ui^I$. They are depicted in
figure \ref{fig:ScalarTadp}. The first and the second tadpole diagrams
result from the first vertex given in figure \ref{fig:Vertices}. The
last two diagrams, both, have a multiplicity of two, since they 
can be obtained from the middle as well as the last vertex of figure
\ref{fig:Vertices}. The expressions for these tadpole diagrams are
given by  
\equ{
\arry{lcl}{
A & = & \dsp \der_i A_\ui^I(z)\, 
\frac 1{3} \int \frac{\d^4 p}{(2\gp)^4} 
~ \sum_{k, \uj, w, q}~
\gf_{qw}^\dag(\gth^{-k} z) \, \frac {1}{p^2 + \gD}\,  \gf_{qw}(z) ~
(~ i f_{Iw}^{~~~w}) ~ 
 \gth^{-\gs(\uj, w)k},
\\[2ex]
B & = & \dsp  ~~~ A_\ui^I(z) \,  
\frac 1{3} \int \frac{\d^4 p}{(2\gp)^4} 
~\, \sum_{k, w, q}~~
\gf_{qw}^\dag(\gth^{-k} z) \, \frac {\der_i}{p^2 + \gD}\,  \gf_{qw}(z)
~  (~ i f_{Iw}^{~~~w}) ~
 \gth^{-\gs(\ui, w)k} (- \bgth^k), 
\\[2ex]
C & = & \dsp  ~~~ A_\ui^I(z) \,  
\frac 1{3} \int \frac{\d^4 p}{(2\gp)^4} 
~\, \sum_{k, w, q} ~~
\gf_{qw}^\dag(\gth^{-k} z) \, \frac {\der_i}{p^2 + \gD}\,  \gf_{qw}(z)
~ ( \mbox{-}i f_{Iw}^{~~~w}) ~
\gth^{-\gs(\ui, w)k} (2), 
\\[2ex]
D & = & \dsp  ~~~ A_\ui^I(z) \,  
\frac 1{3} \int \frac{\d^4 p}{(2\gp)^4} 
~ \sum_{k, \uj, w, q} ~
\gf_{qw}^\dag(\gth^{-k} z) \, \frac {\der_i}{p^2 + \gD}\,  \gf_{qw}(z)
~ (\mbox{-} if_{Iw}^{~~~w}) ~
\gth^{-\gs(\uj, w)k} (2), 
}
}
For diagrams $A$ and $D$ of fig.\ \ref{fig:ScalarTadp}, 
we have to sum over
$\uj$,  since all of the complex components of the internal gauge
field contribute. The last three diagrams of figure \ref{fig:ScalarTadp}
all have $z_i$ derivative inside the loop. It is important to
realize, that only for diagram $B$ the orientation of the
differentiation arrow and the complex index arrow are opposite. This
signifies, that the derivative is not hitting $\gf_{qw}(z)$ but rather 
$\gf^\dag_{qw}(\gth^{-k}z)$. As can be inferred from the first formula 
in \eqref{DerProducts} of appendix \ref{sc:WaveProp}, this implies
that this diagram picks up an additional factor $- \bgth^k$. 

\begin{figure}
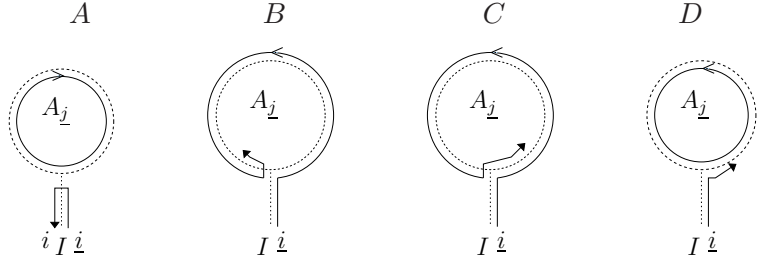

\[
\arry{cccc}{
A & B & C & D 
\\[1ex] 
\raisebox{0ex}{\scalebox{0.8}{\mbox{\input{Tad_FI.pstex_t}}}}
\quad & \quad 
\raisebox{0ex}{\scalebox{0.8}{\mbox{\input{Tad_bosref.pstex_t}}}}
\quad & \quad 
\raisebox{0ex}{\scalebox{0.8}{\mbox{\input{Tad_bos.pstex_t}}}}
\quad & \quad 
\raisebox{0ex}{\scalebox{0.8}{\mbox{\input{Tad_lop.pstex_t}}}}
} 
\]
\caption{The tadpole diagrams for $A_\ui^I$ with internal gauge fields
$A_\uj$ in the loop. The first two diagrams result from the first 
vertex in figure \ref{fig:Vertices}. Each of the last two diagrams 
appears twice because they result from the latter two vertices in 
figure \ref{fig:Vertices}.}
\labl{fig:ScalarTadp} 
\end{figure} 

In addition to these four dimensional scalar loop diagrams, we have
contributions from the four dimensional vector $A_\gm$, the ten
dimensional gaugino $\gch$ and the ghost $\get$. 
As the bulk theory is non--Abelian, the resulting ghost sector does
not decouple from the calculation.  In ten dimensional Feynman
gauge  the ghost has a ten dimensional propagator. In figure 
\ref{fig:RemainTadp} these other tadpole diagrams are collected;  
their contributions read 
\equ{
\arry{lcl}{
E & = & \dsp  ~~~ A_\ui^I(z) \,  
\frac 1{3} \int \frac{\d^4 p}{(2\gp)^4} 
~\sum_{k, w, q,\gm} ~
\gf_{qw}^\dag(\gth^{-k} z) \, \frac {\der_i}{p^2 + \gD}\,  \gf_{qw}(z)
~  (\mbox{-} i f_{Iw}^{~~~w}) 
~ \gth^{-\gs(\gm, w)k}, 
\\[2ex]
F & = & \dsp  \, - A_\ui^I(z) \,  
\frac 1{3} \int \frac{\d^4 p}{(2\gp)^4} 
~ \sum_{k, w, q,\ga} ~ 
\gf_{qw}^\dag(\gth^{-k} z) \, \frac {\der_i}{p^2 + \gD}\,  \gf_{qw}(z)
~  (\mbox{-} i f_{Iw}^{~~~w}) 
~ \gth^{-\gs(\ga\,, w)k}(2), 
\\[2ex]
G & = & \dsp  \, - A_\ui^I(z)\, 
\frac 1{3} \int \frac{\d^4 p}{(2\gp)^4} 
~~ \sum_{k, w, q} ~~
\gf_{qw}^\dag(\gth^{-k} z) \, \frac {\der_i}{p^2 + \gD}\,  \gf_{qw}(z)
~ (\mbox{-}i f_{Iw}^{~~~w})
~ \gth^{-\gs(w)k} (2).
}
\labl{Remain}
}

Before we discuss the details of the evaluation of these tadpoles, we
first turn to the following cancellations, which are ultimately due to 
supersymmetry. Symbolically they may be represented as  
\equ{
E + F_{+++} + G \propto 4 - 2 -2 = 0; 
\qquad
D + F_{(+--)} \propto 3*2 - 3 *2 = 0.
}
Here $F_{+++}$ denotes the expression of $F$ in equation 
\eqref{Remain} (or diagram $F$ of fig.\ \ref{fig:RemainTadp}) with 
the internal gaugino chirality $+++$. Similarly, $F_{(+--)}$ refers to the
sum over the three cyclic permutations of the chirality indices $+--$
of expression $F$. We have used table \ref{tab:SymmFact} to 
conclude that the corresponding expressions are equal up to the given
multiplicities.  
This shows that we are left with the three diagrams
$A, B$ and $C$ (see figure \ref{fig:ScalarTadp}).  
From this point onwards it
is important to distinguish between the cases $k = 0$ and $k \neq 0$;
we will denote the expressions for the corresponding diagrams with
subscripts. Let us first consider the remaining diagrams for $k = 0$.

\begin{figure}
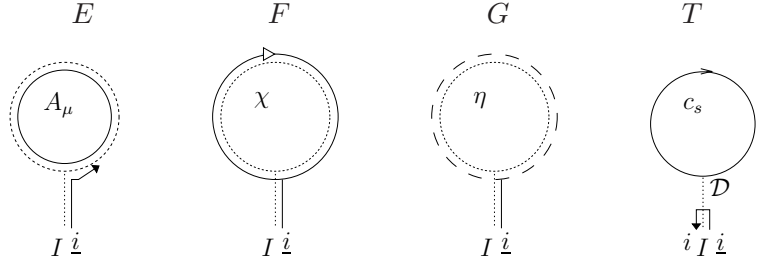

\[
\arry{cccc}{
E & F & G & T
\\[1ex]
\raisebox{0ex}{\scalebox{0.8}{\mbox{\input{Tad_vec.pstex_t}}}}
\quad & \quad 
\raisebox{0ex}{\scalebox{0.8}{\mbox{\input{Tad_fer.pstex_t}}}}
\quad & \quad
\raisebox{0ex}{\scalebox{0.8}{\mbox{\input{Tad_gst.pstex_t}}}}
\quad & \quad 
\raisebox{0ex}{\scalebox{0.8}{\mbox{\input{Tad_twt.pstex_t}}}}
}
\]
\caption{
The remaining tadpole diagrams for $A_\ui^I$ consisting of loops with 
the four dimensional gauge field ($E$), the ten dimensional gaugino ($F$),
the  Fadeev--Poppov ghost ($G$), and the twisted states ($T$).  
}
\labl{fig:RemainTadp}
\end{figure}

\subsubsection*{$\boldsymbol{k = 0}$: bulk tadpoles}

As can be inferred from the discussion below definition
\eqref{OrbiProjc}, the case $k = 0$ corresponds to the calculation on
the torus. It is not difficult to see that $A_{k=0}$ vanishes: The sum is over
all positive and negative $q$ and $w$, by taking $q\ra -q$ and $w \ra
-w$ the resulting expression remains the same, except that the structure
coefficients change sign: 
$f_{I\mbox{-}w}^{~~~\mbox{-}w} = -f_{Iw}^{~~~w}$. But since the
summation indices $q$ and $w$ are dummy indices, this implies that 
$A_{k=0} = - A_{k=0} =0$. For $B_{k = 0}$ and $C_{k=0}$ a similar
argument does not hold: Because of the extra derivative $\der_i$
sandwiched between $\gf_{qw}^\dag$ and $\gf_{qw}$, those expressions
do not vanish. However, the expression for the tadpole can be represented as
a derivative w.r.t.\ $a_\ui^I$  
\equ{
B_{k=0} + C_{k=0} = - i A_\ui^I \, 
\frac { R_i}{ 16 \gp i} \frac {\bgth - \gth}{\gth -1}\, 
\pp[]{a_\ui^I} 
\,  \frac 1{V_6} \sum_{q,w} \int \frac{\d^4 p}{(2\gp)^4} 
\ln \Big[ p^2 + (2\gp)^2 | q_i + b_{i\,w}|^2/R_i^2  \Big]
f_{Iw}^{~~~w},
}
where $V_6$ is the volume of the torus and $b_{i\, w}$ is defined 
in \eqref{Defb} of appendix \ref{sc:WaveProp}. 
This completes the computation of the case $k =0$. 
The interpretation of this tadpole is the following: Because we have
allowed for Wilson lines in the model, it is not surprising, that these
constant gauge backgrounds will receive quantum corrections. 
To see that this interpretation makes sense, we observe that if there
are no Wilson lines: $b_{iw} = 0$. Hence, the whole expression
vanishes due to the derivatives  $\der/\der a_\ui^I$. 
Since our main interest in this paper is to investigate the new
counter terms at one loop, we will ignore this contribution from now on.

\subsubsection*{$\boldsymbol{k \neq 0}$: localized tadpoles}

For the case $k\neq 0$ similar methods can be employed as in the previous
section for the derivation of \eqref{subresultFI}. We need a
subsequent partial integration to put the single derivate $\der_i$  on
the external line of $A_\ui^I$, which gives  
\equ{
\arry{lcl}{
A_{k\neq 0} & = & \dsp \der_i A_\ui^I(z)\, 
\frac 1{3} \int \frac{\d^4 p}{(2\gp)^4} 
\frac {1}{p^2 + \mbox{$\frac 13$} \gD}\, 
\sum_{k, \uj, w, s} 
\frac {\gd(z - \fZ_s -\gG) }{27} ~
(~ i f_{Iw}^{~~~w}) ~
\gth^{-\gs(\uj, w, s)k}, 
\\[2ex]
B_{k \neq 0} & = & \dsp  \der_i  A_\ui^I(z) \,  
\frac 1{3} \int \frac{\d^4 p}{(2\gp)^4} 
\frac {1}{p^2 +  \mbox{$\frac 13$} \gD}\,
~\sum_{k, w, s}~ 
\frac {\gd(z - \fZ_s -\gG) }{27} ~
 (~ i f_{Iw}^{~~~w}) ~
 \gth^{-\gs(\ui, w, s)k} 
 \Bigl( \frac {\bgth^k}{1- \bgth^k} \Bigr) , 
\\[2ex]
C_{k \neq 0}  & = & \dsp  \der_i A_\ui^I(z) \,  
\frac 1{3} \int \frac{\d^4 p}{(2\gp)^4} 
\frac {1}{p^2 +  \mbox{$\frac 13$} \gD} \, 
~\sum_{k, w, s} ~
\frac {\gd(z - \fZ_s -\gG)}{27} ~
(~i f_{Iw}^{~~~w}) ~
\gth^{-\gs(\ui, w, s)k} 
\Bigl( \frac {2}{1- \bgth^k} \Bigr). 
}
}
Here we encounter another, more subtle, cancellation: 
The sum of the contributions $B$ and $C$ is proportional to 
\equ{
\sum_{k = 1,2} \sum_{\rep{r} = \rep{R}_s, \crep{R}_s} 
\frac { 2 + \bgth^k}{1- \bgth^k} \gth^{- \gs(\ui, \rep{r})k} (-)^{\rep{r}}
= - \sum_{k = 1,2} \sum_{\rep{r} = \rep{R}_s, \crep{R}_s} 
 \gth^{- (\gs(\ui, \rep{r}) -1)k} (-)^{\rep{r}} = 0.
}
In the last step we have used that $\gs(\ui, \rep{R}_s) = 0 \mod 3$ 
and $\gs(\ui, \crep{R}_s) = 2 \mod 3$. Therefore, the only
non--vanishing contribution for $k\neq 0$ comes from diagram A of
 figure \ref{fig:ScalarTadp}.  As can be seen from the expression of
$A_{k\neq0}$, the sum over the twist factors is the same as the one 
already computed in \eqref{SymmSurf}. Hence we obtain  
\equ{
A_{k \neq 0} = 
\der_i A_\ui^I \, 
\sum_s 
3 i \, \tr_{\rep{R}_s}(H_I) \, 
\int \frac{\d ^4 p}{(2\gp)^4} 
\frac 1{p^2+ \frac 13 \gD} \, 
\frac 1{27}  \gd(z - \fZ_s - \gG).
}
Again, we use the cut--off scheme to compute the divergent integrals. 
Clearly, the calculation of the tadpole of $A_i^I$ is completely
analogous to the one just presented, except that in the whole calculation
$i \leftrightarrow \ui$. Therefore the two expressions are related by
Hermitian conjugation, and we obtain a relative minus sign. Hence the full
expression for the gauge field tadpoles on the orbifold takes the form:
\equ{
L_{tadp\, un} = 
-\sum_{s}  \, 
3  i\,  \tr_{\rep{R}_s}( \der_i A_\ui - \der_\ui A_i ) \, 
\left( 
\frac {\gL^2}{16 \gp^2} 
+ \frac {\ln \gL^2}{16 \gp^2}\,  
\frac 13 \gD 
\right) 
\frac 1{27}  \gd(z - \fZ_s - \gG).
\labl{full4Dtadp} 
}

In addition to the contributions of the gauge multiplet in the
bulk, the twisted states at the fixed points supply us with an
additional source of a tadpole for $A_\ui^I$, see \eqref{FixPointLagr}. 
As we discussed there, the inclusion of the
twisted states is performed most conveniently using a four dimensional
off--shell formulation. This results in the last tadpole diagram of
figure \ref{fig:RemainTadp} in which the auxiliary field $\cD$ is
exchanged; the tadpole for $A_\ui$ due to twisted states, therefore, 
becomes 
\equ{
L_{tadp\, tw} =  
-\sum_{s}
i \, \tr_{\rep{S}_s + 3\rep{T}_s}(\der_i A_\ui - \der_\ui A_i) \, 
\frac {\gL^2}{16 \gp^2} \, \gd(z - \fZ_s - \gG).
\labl{div4Dtw} 
} 
Hence the total expression for the tadpoles of $A_\ui$ reads 
\equ{
L_{tadp} = -  i \, (\der_i A^I_\ui - \der_\ui A^{I}_{i})
\sum_{s}\Bigl[ 
\tr_{\rep{L}_s}(H_I) \, 
\frac {\gL^2}{16 \gp^2}  + 
\frac 1{27} \tr_{\rep{R}_s}(H_I) 
\frac {\ln \gL^2}{16 \gp^2}\,   \gD 
\Bigr] 
\gd(z - \fZ_s - \gG).
\labl{div4Dtot} 
} 
This is precisely the expression for the localized tadpoles one would
expect on supersymmetry grounds from the tadpole for the auxiliary
field $\cD$, as computed in \eqref{FItadp}. 
An additional cross check of the off--shell $\cD^I$--tadpoles calculated 
in section 4.1, can be provided by a direct computation of the mass terms 
of $A_\ui$. We will not present this in this paper.

\section{Consequences of vanishing $\boldsymbol{\cD}^I$--terms}
\labl{sc:Dterms}

We investigate some consequences of localized FI--tadpoles in
heterotic models. Similar methods will be pursued in this analysis as
those that were used in the five dimensional case of $\U{1}$ gauge
fields on the orbifold $S^1/\Intr_2$ 
\cite{GrootNibbelink:2002wv,GrootNibbelink:2002qp}. 
However, there are various reasons why the analysis in the present
case is in principle more involved: there are more fields in
the game, in particular the gravitational interactions, as well as
the anti--symmetric tensor may be
relevant. Additionally, the (local) Green--Schwarz
mechanism has introduced various interaction terms involving
(non--Abelian) gauge fields.

\subsection{Cartan symmetry breaking}
\labl{sc:breaking}

We turn our attention to one phenomenologically
important issue: spontaneous breaking of gauge symmetries due to 
one--loop induced FI--terms. Since the FI--tadpoles only arise
for the gauge symmetries of the Cartan subalgebra, we investigate 
when spontaneous breaking of the Cartan gauge symmetries is
inevitable. Cartan symmetry breaking occurs, if a field that is
charged under the Cartan subalgebra of $\E{8}\times \E{8}'$ acquires 
a non--vanishing VEV. This can be either an internal gauge field component
$\langle A^w_\ui \rangle $, a twisted state $\langle c_s \rangle$
at a fixed point, or some combination. 
As usual we are looking for a supersymmetric minimum of
the theory from the global four dimensional point of view. 
This means that we can exploit various BPS--like equations, which 
simplify the analysis of the equations of motion considerably. 
Here we do not perform a full analysis, but rather we are only
concerned with the BPS--equations that result from the supersymmetry
transformations of the gauginos in the 
Cartan subalgebra (see \eqref{4DgauginoSusy} 
and \eqref{4DsusyTrans}) 
\equ{
\gd \gch^{+++\, I} 
= - \frac 14 F^{\gm\gn\, I} \gg_{\gm\gn} \ge - \frac i2 \cD^I \tgg \ge, 
}
where we have used the auxiliary field $\cD^I$ to encode the
modifications to the supersymmetry transformation rule due to the
twisted states and the FI--tadpoles. As we are looking for vacuum
configurations that preserve $N=1$ supersymmetry at the global four
dimensional level, i.e.\ $\ge(x,z) = \ge(x)$ is constant over the internal
dimensions, we find that 
\equ{
\langle \cD^I\rangle  = i \langle F^{I}_{i \ui} \rangle  
+ \sum_s \Bigl( \langle \bar c_s \rangle H_I \langle c_s \rangle  
+ \frac {\gL^2}{16 \gp^2}\,  \tr_{\rep{L}_s}(H_I)
+ \frac 1{27} 
\frac {\ln \gL^2}{16 \gp^2}\,  \tr_{\rep{R}_s}(H_I)
 \gD 
\Bigr) \gd(z - \fZ_s - \gG)
= 0.
\labl{CartanBPS} 
}
Here we have assumed that the four dimensional vacuum does not break
Lorentz invariance and hence the VEV of $F^{\gm\gn}$ vanishes. 
In general this equation may be viewed as a BPS equation of motion for
$\langle A_\ui^I\rangle$ and its conjugate, which reside in the field
strength $\langle F^I_{i \ui} \rangle$. 

However, there may not always be a solution of this BPS equation of
motion. If there is no solution, this implies that supersymmetry is
spontaneously broken. Therefore, it is important to investigate
under which condition the BPS equation can be satisfied, and what the
consequences of this condition are. To investigate these questions we
integrate over the extra internal dimensions 
\equ{
\int\limits_{T^6/\Intr_3} \d^6 z \Bigl\{
\sum_s \Bigl( \langle \bar c_s\rangle  H_I \langle c_s \rangle +
\frac {\gL^2}{16 \gp^2}\,  \tr_{\rep{L}_s}(H_I)
+ 
\frac 1{27} 
\frac {\ln \gL^2}{16 \gp^2}\,  \tr_{\rep{R}_s}(H_I)
 \gD 
\Bigr) \gd(z - \fZ_s - \gG)
+ i \langle F^{I}_{i \ui}  \rangle 
\Bigr\} = 0. 
}
Since the orbifold singularities have codimension six, we can simply
use Stoke's theorem to remove the $\der_i A^I_\ui - \der_\ui A^{I}_{i}$
part in the field strength $F_{i \ui}$ and the term proportional
to $\gD$. Hence we are left with the constraint 
\equ{
\sum_s 
\langle \bar c_s \rangle H_I \langle c_s \rangle 
+ w_I 
\int\limits_{T^6/\Intr_3} \d^6 z 
\langle A^{-w}_i\rangle\langle  A_{\ui}^{w} \rangle 
 = 
-\frac {\gL^2}{16 \gp^2} \sum_s \,  \tr_{\rep{L}_s}(H_I). 
\labl{CartanBPSglobal} 
} 
Now, since $\sum_s \tr_{\rep{L}_s}(H_I) H_I$ precisely identifies the
global anomalous $\U{1}$ generator, it follows that if 
$\sum_s \tr_{\rep{L}_s}(H_I) \neq 0$, at least either an internal
gauge field $A_\ui^w$ with $\sum_s  \tr_{\rep{L}_s}(H_I) w_I \neq 0$
or a twisted scalar $c_s$ with $\sum_s \tr_{L_s}(H_I)q_I \ne 0$ has to
get a non--vanishing VEV to  
cancel the FI--tadpole. Observe that the sign of the charge of that
field must be opposite to that of $\sum_s \tr_{\rep{L}_s}(H_I)$. 
This in turn implies that the anomalous $\U{1}$ is spontaneously broken at
the zero mode level. Conversely, if there is no global anomalous
$\U{1}$, i.e.\ if  $\sum_s \tr_{\rep{L}_s}(H_I) = 0$, then
supersymmetry can be preserved without any Cartan symmetry
breaking. (It should be realized that 
other BPS--equations may require that part of the Cartan subalgebra is
spontaneously broken. This is beyond the scope of the present paper.)

\subsubsection*{Anomalous pure orbifold models}

It is instructive to see how Cartan symmetry breaking works 
for the pure orbifold models: both in its own right, and because they
examplify some typical symmetry breaking patterns also to be expected
to appear in models with
Wilson lines. In table \ref{tab:AnomFixedModels} we have given
various characteristics of the two pure orbifold models, $\E{7}$ and
$\SU{9}$: the trace over $\rep{L}_s$, which appears as a factor in the
global (zero mode) FI--term, and the untwisted and twisted matter
representations are given.  Hence we can read of which field(s) may develop
a VEV to cancel the quadratically divergent FI--tadpole. 

In the $\E{7}$ model, there are two types of states which have
negative anomalous charge and therefore would break the anomalous
$\U{1}$: the untwisted state 
$\smash{(\rep{3}_H, (\rep{1})_{\mbox{-}2}(\rep{1})'_{0})}$ 
and the twisted states 
$\smash{(\rep{1}_H,(\rep{1})_{\mbox{-}\frac{4}{3}}(\rep{1})'_{\frac{2}{3}})}$. 
Both types are non--Abelian gauge singlets and will therefore not induce
further spontaneous non--Abelian symmetry breaking. However, in
addition to the charges under the anomalous $\U{1}$, a twisted
singlet is charged under the non--anomalous $\U{1}'$ as well;
therefore its VEV leads to spontaneous breaking of both $\U{1}$'s, the
anomalous and the non--anomalous one. Let us assume that only one type
of states has a non--vanishing VEV and in such a way that the zero
mode FI--term is cancelled. In the $\E{7}$ model the twisted spectra
at all fixed points are the same, therefore it is possible that all
quadratically divergent tadpoles can be canceled locally. However,
since the only condition is the cancellation of the zero mode
(quadratically divergent) tadpole, it might just be one twisted state
at one of the fixed points that cancels the zero mode
tadpole. Likewise the untwisted state  
$(\rep{3}_H, (\rep{1})_{-2}(\rep{1}_0')$ may cancel the quadratically (and 
even the logarithimically) divergent tadpoles locally, but again this is 
not necessary. Observe that contrary to the untwisted states, the twisted 
states can never cancel any of the logarithmic divergences. As these
tadpoles are proportional to the trace $\tr_{\rep{R}_s}(H_I)$ over
the untwisted matter representations $\rep{R}_s$ only, it follows that
both $\U{1}$'s (not just the anomalous one) 
have non--vanishing logarithmically divergent tadpoles, as can be
easily confirmed by consulting the one but last column of table 
\ref{tab:AnomFixedModels}. 

The $\SU{9}$ model has only one representation with a charge opposite
to the quadratically divergent FI--tadpole: the untwisted state
$(\rep{3}_H, (\rep{1})(\rep{14})'_{\mbox{-}1})$. Therefore, like in
the case of the untwisted states in the $\E{7}$ model cancelling the
global Fayet--Iliopoulos tadpole, the shape of the untwisted states of
the $\SU{9}$ model may be such that all localized tadpoles are
cancelled. However, since this state is charged under the $\SO{14}$,
we infer that the model exhibits spontaneous symmetry breaking:
$\SO{14} \ra \SO{13}$. Like for the $\E{7}$ model the $\U{1}'$ has a
non--vanishing logarithmically tadpole
(c.f. table \ref{tab:AnomFixedModels}).

\subsection{Background profile of $\boldsymbol{A_\ui^I}$}
\labl{sc:BackProfile}

From now on we assume that either $\sum_s \tr_{\rep{L}_s}(H_I) = 0$
or that some untwisted states $\langle A_\ui^w \rangle$ and/or twisted
states $\langle c_s\rangle$ have acquired a VEV such that
\eqref{CartanBPSglobal} is satisfied. Then we know that there exists a
solution for $\langle A_\ui^I \rangle$ to \eqref{CartanBPS}. In this
subsection we wish to construct it explicitly. Locally we can
introduce a potential $\langle P^I\rangle$ for 
$\langle A_\ui^I \rangle$ defined by the following equations 
\equ{
\langle A_\ui^I \rangle = i \der_\ui \langle P^I\rangle, 
\qquad 
\langle A_i^I \rangle = -i \der_i \langle P^I\rangle.
\labl{AasP}
}
Substituting this in \eqref{CartanBPS} leads to the equation 
\equ{
\gD \langle P^I\rangle  = 
\sum_s \Bigl( \langle \bar c_s \rangle H_I \langle c_s \rangle  + 
\frac {\gL^2}{16 \gp^2}\,  \tr_{\rep{L}_s}(H_I)
+ 
\frac 1{27} 
\frac {\ln \gL^2}{16 \gp^2}\,  \tr_{\rep{R}_s}(H_I)
 \gD 
\Bigr) \gd(z - \fZ_s - \gG) 
%\\[2ex]
+  w_I\langle A^{-w}_{i}\rangle \langle A^{w}_{\ui} \rangle. 
%\non 
}
To solve this equation, consider first the Green's function $\cG(z-y)$
on $\Cplx^3$ defined by 
\(
\gD_z \cG(z-y) = \gd^6(y).
\)
Since the delta function on the torus is $\gd^6(y - \gG)$, it follows
that the Green's function of the torus $T^6$ reads $\cG(z-y-\gG)$. 
Using this Green's function, it is straightforward to obtain the solution for 
$\langle P^I \rangle$, it reads 
\equ{ 
\arry{rcl}{\dsp 
\langle P^I \rangle &= & \dsp 
\sum_s\Bigl(  \langle \bar c_s \rangle H_I \langle c_s \rangle  + 
\frac {\gL^2}{16 \gp^2}\,  \tr_{\rep{L}_s}(H_I)
\Bigr) \, \cG(z - \fZ_s - \gG) 
\\[2ex]  
&& \dsp + 
\frac 1{27} 
\frac {\ln \gL^2}{16 \gp^2} 
\sum_s  
\tr_{\rep{R}_s}(H_I) \, \gd(z - \fZ_s - \gG)
\\[2ex]  
&& \dsp 
+ w_I\, 
\int_{T^6/\Intr_3} \d^6y\, 
\langle A^{-w}_{i}(y) \rangle \langle A^{w}_{\ui}(y) \rangle \, 
\cG(z - y - \gG).
}
}
Let us make a couple of comments:
It might seem that it always provides us with a solution. But that is
only a local statement, which ignores the crucial compactness that
resulted in the constraint \eqref{CartanBPSglobal}. 

If a bulk state $A_\ui^w$ is required to get a VEV to satisfy the
global BPS condition \eqref{CartanBPSglobal}, its shape over the extra
dimensions might be quite complicated. (It will not be determined in
this paper.) However, whatever precisely its profile is, this formula
gives the resulting shape of $\langle A_\ui^I \rangle$. 

Suppose another situation, where all anomalous fixed points are
equivalent to an $\E{7}$ model, then the discussion of the previous
subsection tells us that it is possible to cancel all quadratically
divergent tadpoles locally by giving VEVs to the twisted singlets 
$\smash{(\rep{1}_H,(\rep{1})_{\mbox{-}\frac{4}{3}}(\rep{1})'_{\frac{2}{3}})}$.
This means that the first and the final line are zero, but the middle
line will still be present. But still, as observed in the previous
section, at the anomalous fixed points the logarithmically divergent
tadpoles are present. From the analysis here we infer that they lead
to the profile 
\equ{
\langle A_\ui^I \rangle = 
\frac i{27} 
\frac {\ln \gL^2}{16 \gp^2} 
\sum_s  
\tr_{\rep{R}_s}(H_I) \, \der_\ui \, \gd(z - \fZ_s - \gG),
}
with a derivative of the fixed point delta function.

\section{Conclusions}
\labl{sc:concl}

In this paper we investigated the role of localized anomalous
$\U{1}$'s, which appear at the fixed points of heterotic orbifold
compactifications with Wilson lines, for the case $T^6/\Intr_3$. The
main results of this work are summarized as follows:

The first question we addressed was how gauge invariance at the
fixed points (with anomalous
$\U{1}$'s) is restored. We showed that by using a local version of
the Green--Schwarz mechanism at the fixed points, the localized
pure and mixed $\U{1}$ anomalies, that arise due to the ten
dimensional gauginos, are canceled. Also 1/3 of the original ten
dimensional anomaly is present on the orbifold.  
We checked explicitly that the ten
dimensional and the four dimensional local fixed point Green--Schwarz
mechanisms are compatible with each other.

Next we investigated whether these localized anomalous $\U{1}$'s are
associated with Fayet--Iliopoulos tadpoles, as is the case for the
well--know situation at the zero mode level. To this end it proved
useful to construct an off--shell formulation of the full ten
dimensional super Yang--Mills theory with respect to the four
dimensional $N=1$ supersymmetry, which is left unbroken by the
orbifolding. The Fayet--Iliopoulos 
tadpole diagrams with gauge fields in the loop were computed with the
help of the
orbifold projector method. Using cut--off regularization, we found
that the quadratically divergent part of these tadpoles are
proportional to the same traces as the anomalous $\U{1}$'s (twisted
states taken into account). However, we found also logarithmically
divergent terms, which scale with the traces over the untwisted
states only, and appear together with the double derivative of the
orbifold delta functions. (These results are very similar to the ones obtained
previously for five dimensional orbifold models 
\cite{
Barbieri:2002ic,GrootNibbelink:2002wv,GrootNibbelink:2002qp,Scrucca:2001eb}.) 
Because of supersymmetry, one would expect similar tadpoles to arise
for the 
internal part of the Cartan gauge field strengths. We confirm this by 
a direct (on--shell) calculation of the tadpoles of the internal gauge
fields.

In the final part of this article we investigated some 
consequences of such tadpoles. First we studied the BPS equations
$\cD^I=0$, which are required for unbroken $N=1$ supersymmetry in four
dimensions. 
We found that they can only be solved, if the global Fayet--Iliopoulos
tadpole is canceled by a VEV of at least one charged bulk or twisted
field. In this 
way we rederived the standard four dimensional zero mode conditions
for unbroken $N=1$ supersymmetry, and spontaneous breaking of the zero
mode anomalous $\U{1}$. However, it turned out that these global BPS
conditions 
can be solved in numerous ways, corresponding to different profiles
of the charged untwisted states over the orbifold and different VEVs
of the charged twisted states. (We have discussed some of these 
possibilities for
the pure orbifold models $\E{7}$ and $\SU{9}$, that are defined in table
\ref{tab:AnomFixedModels}.) The background profiles of the internal
Cartan gauge fields $A_\ui^I$ have been determined in general, using
the only constraint, that the global BPS condition was satisfied.

\subsection*{Outlook}

We would like to make a couple of remarks on further developments along
the lines of the present work.

First of all we should stress that all the results presented in this
paper were obtained using pure field theory arguments. But since we
are really describing the low energy limit of heterotic string
theory, it would be interesting to see if our results can be
confirmed by full string computations as well. In particular the
localized version of the Green--Schwarz mechanism, and the structure
of the tadpoles of the internal field strength may also be calculated
using string theory techniques.

As compared to the five dimensional situation, one may wonder whether
also (strong) localization effects of untwisted states can arise. (For
the localization effects in five dimensional orbifold models see 
\cite{GrootNibbelink:2002wv,GrootNibbelink:2002qp,Marti:2002ar,Abe:2002ps}).  
In this work we have obtained the profile of the background of the bulk
gauge field. Therefore, in principle localization effects can be
investigated. However, there is one technical hurdle to overcome here:
not only the gauge connection, but also the spin connection appears in
the equation of motion of the gaugino. Because of the curvature
singularities, this spin connection is also strongly peaked at the
orbifold fixed points. It is therefore questionable, whether it
suffices to take only the gauge connection background into account to
investigate the possibility of localization effects.

Finally, the methods and the results obtained in this paper for the
$\Intr_N$ orbifold with $N=3$, can be generalized for higher
$N$. Especially for non--prime $N$ it may be interesting to see 
what kind of tadpoles can arise and to compare the results with localized
anomalies on such orbifolds.

\section*{Acknowlegdments}

We would like to thank M.\ Laidlaw and E.\ Poppitz for discussions.

SGN acknowledges the support of CITA and NSERC. The work of HPN 
and MW has been supported in part by the European Community's Human
Potential Programme under contracts HPRN--CT--2000--00131 Quantum  
Spacetime, HPRN--CT--2000--00148 Physics Across the Present Energy
Frontier and HPRN--CT--2000--00152 Supersymmetry and the Early
Universe. The latter Programme also supported MO.

\appendix

\section{Decomposition of ten dimensional spinors}
\labl{sc:Spinor6D}

This appendix provides some useful properties of ten dimensional
spinors and their decomposition to a four dimensional Minkowski space
times an three dimensional complex internal manifold. (Details can be
found in \cite{VanProeyen:1999ni,pol_2}.) To change from the six
dimensional real coordinates, \{$x^4, \ldots, x^9$\}, to the three
dimensional complex coordinates, \{$z_1, z_2, z_3$\},  we use the 
following redefinitions  
\equ{
x^{2i+2} = x_{2i+2}  = \frac 1{\sqrt{2}} ( z_i +  \bz_\ui ), 
\qquad 
x^{2i+3} = x_{2i+3} = \frac {-i}{\sqrt{2}} ( z_i -  \bz_\ui ), 
\labl{CmplxBss}
}
and the induced transformations on covariant vectors. Here, we have
used that we work with a metric with the signature:  
$\text{diag}(-1, 1^9)$. Hence for the six dimensional part of the
ten dimensional Clifford algebra we get 
\equ{
\arry{c}{
 \gG^{2i+2} =  \gG_{2i+2} =  \frac 1{\sqrt{2}} ( \gG_i + \gG_\ui ),
\\[1ex]
 \gG^{2i+3} = \gG_{2i+3} = \frac {-i}{\sqrt{2}} ( \gG_i - \gG_\ui ), 
}
\qquad 
\{ \gG_i, \gG_\uj \} = 2 \gd_{i\uj}. 
}
For the decomposition of the ten dimensional supersymmetry transformations 
in section \ref{sc:4Dsusy} it will be convenient to rewrite this
six dimensional internal Clifford algebra in terms of two dimensional
Clifford algebras. 

A convenient complex basis for the two dimensional  Clifford algebra
is defined by  
\equ{
\gs_+ = \pmtrx{0 & \sqrt 2  \\ 0 & 0}, 
\quad  
\gs_- = \pmtrx{0 & 0 \\ \sqrt 2 & 0}, 
\quad 
\gs_3 = \pmtrx{ 1 & 0 \\ 0 & -1}, 
\quad
\gs_0 = \pmtrx{1 & 0 \\ 0 & 1}, 
}
where $\gs_\pm =   ( \gs_1 \pm i \gs_2)/ {\sqrt{2}}$. It has the properties 
\equ{
\gs_3 \gs_\pm = - \gs_\pm \gs_3 = \pm \gs_\pm, 
\quad
\gs_\pm^2 = 0, 
\quad 
\gs_+ \gs_- = 2 \gp_+ = 2 \pmtrx{1 & 0 \\ 0 & 0},
\quad 
\gs_- \gs_+ = 2 \gp_- = 2 \pmtrx{0 & 0 \\ 0 & 1}.
}
Let $\get_\gk$ with $\gk = \pm$ form the basis of two dimensional
spinors, with the properties 
\equ{
\gs_0 \get^\gk = \get^\gk, 
\quad 
\gs_3  \get^\gk = \gk \get^\gk,
\quad 
\gs_\pm \get^\mp = \sqrt 2\, \get^\pm, 
\quad 
\gs_\pm \get^\pm = 0,
\quad
{\get^\gk}^\dag \get^{\gk'} = \gd^{\gk \gk'}. 
\labl{BasicCliffordAction}
}
By introducing the notation 
\(
\gs_{\ga_3\ga_2\ga_1} = 
\gs_{\ga_3} \otimes \gs_{\ga_2} \otimes \gs_{\ga_1}, 
\)
with $\ga_i = 0, \pm, 3$, we can represent the six dimensional
Clifford algebra in complex coordinates (here and below the tensor
products are understood). 

The embedding of this six dimensional Clifford algebra
in the ten dimensional one can then be represented as   
\equ{
\gG_\gm = \Id_6 \gg_\gm, 
\quad
\arry{ccc}{
\gG_1 = \gs_{00+}\tgg,
&
\gG_2 = \gs_{0+3}\tgg,
&
\gG_3 = \gs_{+33}\tgg,
\\ 
\gG_{\ubar 1} = \gs_{00-}\tgg,
&
\gG_{\ubar 2} = \gs_{0-3}\tgg,
&
\gG_{\ubar 3} = \gs_{-33}\tgg,
}
}
where $\tgg$ is the four dimensional chirality operator. The
(anti--symmetric) products of the six dimensional Clifford algebra
generators have been collected in  table \ref{tab:CmplxClifford}. The
main advantage of this basis is that the action of the Clifford
algebra elements on the six dimensional spinors $\get_{\gk_3 \gk_2
  \gk_1}$ can be worked out straightforwardly. 

\begin{table}[h!]
\equ{
\renewcommand{\arraystretch}{1.5}
\arry{r | lll}{ 
\gG_i, ~ \gG_\ui   & 
\gG_1 = \gs_{00+}\tgg
&
\gG_2 = \gs_{0+3}\tgg
&
\gG_3 = \gs_{+33}\tgg
\\ 
 & 
\gG_{\ubar 1} = \gs_{00-}\tgg
&
\gG_{\ubar 2} = \gs_{0-3}\tgg
&
\gG_{\ubar 3} = \gs_{-33}\tgg
\\ \hline 
\gG_{ij}  & 
\gG_{12} = - \gs_{0++}
&
\gG_{23} = - \gs_{++0}
&
\gG_{13} = - \gs_{+3+}
\\ \hline 
\gG_{i\uj}  &
\gG_{1\ubar 1} = \ \ \gs_{003} 
& 
\gG_{1\ubar 2}= - \gs_{0-+} 
&
\gG_{1\ubar 3}= - \gs_{-3+} 
\\ &
\gG_{2\ubar 1} = -\gs_{0+-} 
& 
\gG_{2\ubar 2} = \ \  \gs_{030} 
& 
\gG_{2\ubar 3} = -\gs_{-+0} 
\\ & 
\gG_{3\ubar 1} = -\gs_{+3-} 
& 
\gG_{3\ubar 2} = -\gs_{+-0} 
& 
\gG_{3\ubar 3} = \ \ \gs_{300}
\\ \hline 
\gG_{ijk}  & 
\gG_{123} = - \gs_{+++} \tgg 
\\ \hline 
\gG_{ij\uk}  & 
\gG_{12\ubar 1} = - \gs_{0+0} \tgg 
& 
\gG_{13\ubar 1} = - \gs_{+30} \tgg 
&
\gG_{23\ubar 1} = - \gs_{++-} \tgg 
\\ & 
\gG_{12\ubar 2} = \ \  \gs_{03+} \tgg 
&
\gG_{13\ubar 2} =-\gs_{+-+} \tgg 
&
\gG_{23\ubar 2} =-  \gs_{+03} \tgg 
\\ &
\gG_{12\ubar 3}   =  -\gs_{-++} \tgg 
& 
\gG_{13\ubar 3} =  \ \ \gs_{30+} \tgg 
&
\gG_{23\ubar 3} = \ \ \gs_{3+3} \tgg 
\\ \hline 
\gG_{123\ui}  &
\gG_{123\ubar 1} ~~~ = - \gs_{++3} 
& 
\gG_{123\ubar 2} ~\,  =  \ \ \gs_{+0+} 
& 
\gG_{123\ubar 3} ~\,  = -  \gs_{3++} 
\\ \hline
\gG_{ij\uk\ul}  &
\gG_{12\ubar 1\, \ubar 2} ~~\, = -  \gs_{033} 
& 
\gG_{13\ubar 1\, \ubar 2} ~ = \ \  \gs_{+-3} 
& 
\gG_{23\ubar 1\, \ubar 2} ~ = -  \gs_{+0-} 
\\  & 
\gG_{12\ubar 1\, \ubar 3} ~~\, = \ \ \gs_{-+3} 
&
\gG_{13\ubar 1\, \ubar 3} ~ = -\gs_{303} 
&
\gG_{23\ubar 1\, \ubar 3} ~ = \ \ \gs_{3+-} 
\\ & 
\gG_{12\ubar 2\,  \ubar 3} ~~\, = - \gs_{-0+} 
& 
\gG_{13\ubar 2\, \ubar 3} ~ = \ \ \gs_{3-+} 
& 
\gG_{23\ubar 2\, \ubar 3} ~ = - \gs_{330} 
\\ \hline 
\gG_{123\ui\uj}  & 
\gG_{123\ubar 1 \, \ubar 2} ~\,   = - \gs_{+00} \tgg
& 
\gG_{123\ubar 1 \, \ubar 3} = \ \ \gs_{3+0} \tgg
& 
\gG_{123\ubar 2 \, \ubar 3} = - \gs_{33+} \tgg
\\ \hline 
\gG_{123\ubar 1 \, \ubar 2\, \ubar 3}  & 
\gG_{123\ubar 1 \, \ubar 2 \,\ubar 3} = - \gs_{333}
}
\non}
\caption{The complete basis for the six dimensional internal Clifford
algebra within the 10 dimensional Clifford algebra is given, up to 
Hermitian conjugation. 
}
\labl{tab:CmplxClifford}
\end{table}

Using the six dimensional  $\get_{\gk_3 \gk_2 \gk_1}$, a ten
dimensional Majorana--Weyl spinor $\gch$ can be decomposed in terms of four
dimensional spinors $\hat\gch^{\gk_3 \gk_2 \gk_1}$ as 
\equ{
\gch = \sum_\gk 
\get_{\gk_3 \gk_2 \gk_1} \hat\gch^{\gk_3 \gk_2 \gk_1}.
}
The Majorana--Weyl conditions then lead to the following relations on
the four dimensional spinors: 
\equ{
\gk_1 \gk_2 \gk_3 \, \tgg\,  \hat \gch^{\gk_3 \gk_2 \gk_1} = 
(- \gk_1 \gk_3)  
( \hat\gch^{\mbox{-}\gk_3 \mbox{-}\gk_2 \mbox{-}\gk_1})^{C_-} = 
\hat \gch^{\gk_3  \gk_2 \gk_1}.
\labl{MWDecomp}
}
We can define a basis of four Majorana fermions in $D=(1,3)$
dimensions: $\gch^{\gk_1\gk_2\gk_3}$ with $ \gk_3 \gk_2 \gk_1 = 1$,
such that 
\equ{
\gch^{\gk_3\gk_2\gk_1}_L = (\gk_1 \gk_3)\, 
\hat\gch^{\gk_3\gk_2\gk_1}
 \qquad 
\gch^{\gk_3\gk_2\gk_1}_R  = (\gk_1 \gk_3)\, 
(\hat\gch^{\gk_3\gk_2\gk_1})^{C_-}
}
The expansion of the ten dimensional spinor then takes the form 
\equ{
\gch = \sum_{\gk_1\gk_2\gk_3=+} 
(\gk_1 \gk_3) 
\Bigl( 
\get_{\gk_3 \gk_2 \gk_1} \gch_L^{\gk_3 \gk_2 \gk_1} 
- \gk_2 \, 
\get_{\mbox{-}\gk_3 \mbox{-}\gk_2 \mbox{-}\gk_1} 
\gch_R^{\gk_3 \gk_2 \gk_1} 
\Bigr). 
\labl{Expn10DF}
}
The factor $(\gk_1 \gk_3)$ has been included for notational
convenience: it ensures that the signs appearing in
\eqref{4DchiralSusy} are all the same.

\section{Anomaly polynomials and factorization} 
\labl{sc:AnomPolyFact}

It is well known, that the anomaly is determined by the Wess--Zumino
consistency condition \cite{Wess:1971yu} up to an overall
normalization factor. The solution to this consistency condition can
be obtained from the  characteristic class whose integral gives the
index of the Dirac operator in two dimensions higher
\cite{Nakahara:1990th}.  Let $F_2$ be the curvature 2--form 
of a connection $A_1$ (for example a Yang--Mills gauge connection 
and/or a spin connection), and $\tilde I$ any analytic function. 
By taking the trace and restricting to a $2n+2$--form, we obtain a
closed and invariant form $I_{2n+2}$ defined as 
\equ{ 
\tilde I_{2n+2}(F_2) = \tilde I (F_2) \Big|_{2n+2}, 
\quad 
I_{2n+2}(F_2) = \tr \tilde I_{2n+2}(F_2) 
\quad 
\d I_{2n+2}(F_2) = 0, 
\quad 
\gd_\gL I_{2n+2}(F_2) = 0,
}
which exists locally because of Poincar\'e's lemma. 
Here $\gL$ represents a gauge or local Lorentz transformation. 
Using the descent equations 
\equ{
I_{2n+2}(F_2) = \d I_{2n+1}(A_1), 
\qquad 
\gd_{\gL} I_{2n+1}(A_1) = \d I_{2n}^1(\gL, A_1),
\labl{Descent}
}
the $2n+1$ and $2n$ forms, $I_{2n+1}$ and $I_{2n}^1$ can be determined
explicitly. For example $I_{2n}^1$ is given by the integral
expression 
\equ{
I_{2n}^1(\gL, A) = \Bigl( \frac i{2\pi} \Bigr)^{2} 
\int_0^1 d t\, (1-t) \text{str} \Bigr[
\gL\, \d \, \Bigl\{ A \tilde I_{2n-2}( t\, \d A + t^2 \, A^2 ) \Bigr\}
\Bigr].
\labl{GenAnomaly}
}
Of course, all these anomaly polynomials are still dependent on the
chiral matter content under consideration. This is
encoded in the trace $\tr$. ($\text{str}$ denotes the fully symmetrized
trace.) 

The Green--Schwarz mechanism relies on the factorization of 
anomaly polynomials. This special property can be stated as 
\equ{
I_{2p+2q+4}(F_2) = b_{p,q}\,  I_{2p+2}(F_2) I_{2q+2}(F_2), 
\labl{Factorization}
}
for some integers $p,q \geq 0$, and a representation dependent
proportionality factor $b_{p,q}$. When applying the descent relations
on the product of two anomaly polynomials, we find an additional free
parameter $\ga_{p,q}$ because 
\equ{
\d \Bigl( \ga_{p,q}  I_{2p+1}(A_1)\,  I_{2q+2}(F_2)  +  (1-\ga_{p,q}) 
I_{2p+2}(F_2)\,  I_{2q+1}(A_1) \Bigr) =  I_{2p+2}(F_2)\,  I_{2q+2}(F_2). 
\labl{FactDiff}
}
This seems to lead to an ambiguity in the definition of $I^1_{2n}$. 
However, as has been  discussed in \cite{Green:1985bx}, the constant
$\ga_{p,q}$ is in fact fixed by the following observations. 

The form of the anomaly, given in \eqref{GenAnomaly}, 
applies for any (gauge) connection $A_1$. The factorization relies on
the underlying property of the trace of the anomaly polynomials:   
\equ{
\tr [ T_{1} \ldots T_{p+1} \, T_{p+2} T_{p+q+2} ] = 
 c_{p,q} \,  \tr [ T_{1} ... T_{p+1} ] \, \tr [ T_{p+2} ... T_{p+q+2} ] + 
\text{cyclic perm}. 
\labl{TraceFact}
}
Since the trace is cyclic, we have to account for all cyclic
permutations on the right hand side. As we only want to determine the 
constant $\ga_{p,q}$, we can safely restrict ourselves to the Abelian 
situation and apply this relation directly to the anomaly
\eqref{GenAnomaly}. In this way, we obtain the expression 
\equ{
I^1_{2p+2q+2}(\gL, A_1) = c_{p,q} \, \frac {(p\! +\! 1)! (q\! +\!
1)!}{(p\! +\! q\! +\! 2)!} 
\Bigl( 
(p\! +\! 1)\, I_{2p}^1(\gL, A_1) \, I_{2q+2}(F_2) 
+ (q\! +\! 1)\, I_{2p+2}(F_2) \,I_{2q}^1(\gL, A_1)  
\Bigr),
}
by performing the integral over the variable $t$. Comparing this with
the expressions \eqref{Factorization}, \eqref{FactDiff} and \eqref{TraceFact}
given above, we conclude that 
that 
\equ{
b_{p,q} = c_{p,q} \, \frac {(p+1)! (q+1)!}{(p+q+1)!}, 
\ \text{and} \ 
\ga_{p,q}  =  \frac {p+1}{p+q+2}.
}
Since there was just one parameter ($\ga_{p,q}$) to be fixed in the Abelian
as well as the non--Abelian case, it follows that these results hold
in general. Therefore, we obtain that if an anomaly polynomial 
factorizes like \eqref{Factorization}, then the anomaly takes the form 
\equ{
I^1_{2p+2q+2}(\gL, A_1) = b_{p,q} \,
\Bigl(
\frac{p\! +\! 1}{p\! +\! q \! +\! 2}
\, I_{2p}^1(\gL, A_1) \, I_{2q+2}(F_2) 
+ \frac{q\! +\! 1}{p\!+\! q \!+\! 2}
\, I_{2p+2}(F_2) \,I_{2q}^1(\gL, A_1)  
\Bigr). 
\labl{FactAnom} 
}

\section{trace decompositions of $\E{8}$}
\labl{sc:E8traces}

In this appendix we verify, that for both equivalent models with an
anomalous $\U{1}$ (the $\E{7}$ and $\SU{9}$ models discussed section
\ref{sc:AnomModels}), the following relation is valid  
\equ{
\frac 1{60} \left. \tr_{\E{8}\!\times\! \E{8}'} F^2 \right|_s
= \sum_a \tr F^2_{(a)}
}
when we restrict the quadratic $\E{8} \times \E{8}'$ trace to gauge
group $G_s$ of one of those anomalous models. (That is 
$G_s = \E{7} \!\times\! \U{1} \times \SO{14}' \!\times\! \U{1}'$ or 
$G_s = \SU{9}  \times \SO{14}' \!\times\! \U{1}'$.) The sum $a$ is 
over the gauge group factors in $G_s$. Here we have defined 
\equ{
\tr F^2_{(a)} = \frac 1{I_2^a} \tr_{\rep{fund}} F^2_{(a)},
\quad 
\tr F_{\U{1}}^2 = 2 F_{\U{1}}^2, 
\quad 
\tr F_{\U{1}'}^2 = 4 F_{\U{1}'}^2, 
}
for the non--Abelian $G_{(a)}$ and Abelian group factors, respectively. 
$I_2^a$ denotes the quadratic indices for the
non--Abelian group factors, given in table \ref{tab:indices}. The
normalizations for the $\U{1}$s stem from the levels $k_{q_s} = 2$ 
and $k_{q'_s} = 4$, which we get in our conventions.
  
The remainder of the appendix is composed as follows: in appendix 
\ref{sc:char} a number of useful features of characters are
reviewed. These properties are then used in
the subsequent subappendices to compute the quadratic traces
for the gauge groups that appear in the anomalous $\U{1}$ models.

\subsection{General properties of characters}
\labl{sc:char}

To relate traces of the field strength $i F$ in various
representations of different groups to each other, a convenient tool
is the Chern character 
\equ{
\text{ch}_{\rep{r}}[ i F]_{\rep{Ad}} = 
\tr_{\rep{r}} \exp \Bigl[ \frac {i F}{2\gp} \Bigr]_{\rep{Ad}}. 
}
Here $\rep{Ad}$ denotes the algebra in which the field strength $iF$
lives, and $\rep{r}$ denotes the representation of this algebra over
which the trace is taken. From the definition of the character it
follows,  that the dimension of a representation is given by 
\(
|\rep{r}|  = \dim \rep{r} = \text{ch}_{\rep{r}}[0]_{\rep{Ad}} 
= \tr_{\rep{r}}[\Id]_{\rep{Ad}}. 
\)
(Many useful properties of characters and indices are
collected in \cite{vanRitbergen:1998pn}.) 
The following properties of the Chern character are very useful
\equ{
\text{ch}_{\rep{r}_1 \times \rep{r}_2}[ i F]_{\rep{Ad}} = 
\text{ch}_{\rep{r}_1}[ i F]_{\rep{Ad}} \cdot 
\text{ch}_{\rep{r}_2}[ i F]_{\rep{Ad}}, 
\qquad 
\text{ch}_{\rep{r}_1 + \rep{r}_2}[ i F]_{\rep{Ad}} = 
\text{ch}_{\rep{r}_1}[ i F]_{\rep{Ad}} + 
\text{ch}_{\rep{r}_2}[ i F]_{\rep{Ad}}.  
}
For example, the trace over the adjoint $\rep{Ad}_N = \rep{N^2-1}$ 
of $\SU{N}$ over the field strength squared, can be expressed as:
\equ{
\rep{Ad}_N + \rep{1} = \rep{N} \times \crep{N} 
\quad \Ra \quad
\tr_{\rep{Ad}_N} [ (i F)^2]_{\rep{Ad}} 
= 2N\, \tr_{\rep{N}} [ (i F)^2 ]_{\rep{Ad}}. 
\labl{TrAdSUN}
}
Next we obtain the characters for anti--symmetric representations 
obtained from a representation $\rep{r}$. We denote the $k$th
totally anti--symmetric product of $\rep{r}$ by $[\rep{r}]_k$. (Of
course, we set $[\rep{r}]_0 = \rep{1}$ and $[\rep{r}]_1 = \rep{r}$.) 
Because the determinant, in the representation $\rep{r}$, is fully
anti--symmetric, we can define the generating function of the
characters of the anti--symmetric products $[\rep{r}]_k$ as 
\equ{
\sum_{k=1}^{|\rep{r}|}
x^k \, \ch{[\rep{r}]_k}{iF}
= \det{}_{\rep{r}} \Bigl( 1 + x e^{iF} \Bigr) 
= \exp G_{\rep{r}} (x, iF). 
}
The function $G_{\rep{r}}(x, iF)$ has the properties:
\equ{
G_{\rep{r}}(x, iF) = \sum_{n\geq 1} \frac {(-)^{n-1}}{n} x^n
\ch{\rep{r}}{n\, iF},  
\qquad 
G^p_{\rep{r}} = \left. \Bigl( \frac {\der}{\der x} \Bigr)^p 
G_{\rep{r}}(x, iF) \right|_{x=0} = 
(-)^{p-1} (p-1)! \ch{\rep{r}}{p\, iF}.
}
In the following we only need the first (non--trivial) characters 
of fully anti--symmetric representations $[\rep{r}]_k$ explicitly:
\equ{
\arry{l}{
\ch{[\rep{r}]_2}{iF} = 
\frac 12 \Bigl[ (\ch{\rep{r}}{iF})^2 - \ch{\rep{r}}{2iF} \Bigr], 
\\[2ex]
\ch{[\rep{r}]_3}{iF} = 
\frac 1{3!} \Bigl[ (\ch{\rep{r}}{iF})^3 
-3 \ch{\rep{r}}{iF} \ch{\rep{r}}{2iF}
+2 \ch{\rep{r}}{3iF}
 \Bigr], 
\\[2ex]
\ch{[\rep{r}]_4}{iF} = 
\frac 1{4!} \Bigl[ (\ch{\rep{r}}{iF})^4 
-6 (\ch{\rep{r}}{iF})^2 \ch{\rep{r}}{2iF}
+8 \ch{\rep{r}}{iF}  \ch{\rep{r}}{3iF}
\Bigr. 
\\ \Bigl. 
+3 (\ch{\rep{r}}{2iF} )^2 
-3! \ch{\rep{r}}{4iF}
 \Bigr]. 
}
}
By substituting $iF =0$ these characters give the dimensions of the
representations: 
\equ{
\dim [\rep{r}]_2 = \frac {|\rep{r}|-1}{2} |\rep{r}|, 
\quad 
\dim [\rep{r}]_3 = \frac {|\rep{r}|^2 - 3|\rep{r}| +2 }{3!} |\rep{r}|,
\quad 
\dim [\rep{r}]_4 = \frac {|\rep{r}|^3 - 6|\rep{r}|^2 +11|\rep{r}| -6 }{4!} |\rep{r}|.
\labl{DimAntiSym}
}
Furthermore, for a simple Lie group the traces of $(iF)^2$ over these
anti--symmetric representations read 
\equ{
\arry{c}{
\tr_{[\rep{r}]_2} (iF)^2 = ( |\rep{r}| - 2) \tr_{\rep{r}}(iF)^2, 
\qquad 
\tr_{[\rep{r}]_3} (iF)^2 = 
\frac{|\rep{r}|^2 - 5 |\rep{r}| + 6}{2} 
\tr_{\rep{r}}(iF)^2, 
\\[2ex]
\tr_{[\rep{r}]_4} (iF)^2 = 
\frac{|\rep{r}|^3 - 9 |\rep{r}|^2   +26 |\rep{r}| - 24}{6} 
\tr_{\rep{r}}(iF)^2.
}
}
It is important to note that these formulae can be applied for any
representation $\rep{r}$ not necessarily the fundamental one. 

\begin{table}
\[
\arry{|l|c|c|}{
\hline 
\text{group} & \text{fund.\ repr.} & \text{quadr.\ index} 
\\ \hline 
\E{8} & 248 & 60 \\ \hline 
\E{7} & 56 & 12 \\ \hline 
\SO{14} & 14 & 2 \\ \hline 
\SU{9} & 9 & 1 \\ \hline 
}
\]
\caption{The relevant quadratic indices of the fundamental
representations of the (simple) gauge groups 
that arise in the models with an anomalous $\U{1}$.}
\labl{tab:indices}
\end{table}

\subsection{Quadratic traces of the anomalous fixed point models}
\labl{sc:QuadrTr} 

The quadratic indices and reference representations of the relevant
groups are given in
table \ref{tab:indices}. It is conventional to normalize the indices
w.r.t.\ the fundamental  representation of $\SU{n}$. We now compute
$(1/60)\Tr (iF)^2|_{G_s}$ where $G_s$ is the local gauge group of  
one of the two local anomalous equivalent models: 
$G_s = \E{7}\!\times \!\U{1} \times \SO{14}'\!\times\!\U{1}'$ and 
$G_s = \SU{9} \times \SO{14}'\!\times\! \U{1}'$. 

\subsubsection*{$\boldsymbol{\E{7}}$ quadratic trace}

From the branching rules \eqref{branchings} we see that we 
obtain two times the reference representation $\rep{56}$ and once 
the adjoint representation $\rep{133}$ of $\E{7}$. To relate 
the traces of these two representations, we use their decompositions 
under the branching 
\equ{
\E{7}  \ra  \SU{8}: 
\qquad 
\rep{56}  \ra  \rep{28} + \crep{28}, 
\quad 
\rep{133}  \ra  \crep{70} + \rep{63}.
}
To be able to use the general formulae derived above, we identify these
representations as follows: $\rep{63} = \rep{Ad}_8$, 
$\rep{28} = [\rep{8}]_2$ and $\rep{70} = [\rep{8}]_4$. (Using the
dimension formulae \eqref{DimAntiSym} for the anti--symmetrized
representations these
identifications can be confirmed easily.) Moreover, for the quadratic
traces we find  
\equ{
\tr_{\rep{56}} (iF)^2 = 12\, \tr_{\rep{8}} (iF)^2, 
\qquad 
\tr_{\rep{133}} (iF)^2 = 36\, \tr_{\rep{8}} (iF)^2.
} 
(This confirms that the quadratic index of $\E{7}$ equals $12$.) We
conclude that  
\equ{
\frac 1{60} \tr_{\rep{248}}[(iF)^2]_{\E{7}} = 
\frac 1{12} \tr_{\rep{56}}[(iF)^2]_{\E{7}}.  
}

\subsubsection*{$\boldsymbol{\U{1}}$ quadratic trace} 

For the $\U{1}$ factor in the first $\E{8}$ the computation 
is more straightforward, since we only have to take the dimensions of 
the $\E{7}$ representations and their charges into account. This gives 
\equ{
\frac 1{60}\tr_{\rep{248}}[(iF)^2]_{\U{1}} = 
\frac 1{60} \Bigl( 2 \cdot 56 \cdot (\pm 1)^2 + 2\cdot (\pm 2)^2 \Bigr) 
[(iF)^2]_{\U{1}} = 2 [(iF)^2]_{\U{1}}. 
}

\subsubsection*{$\boldsymbol{\SU{9}}$ quadratic trace} 

For the quadratic trace in the adjoint of $\E{8}$ the relevant branching 
is given in \eqref{branchings}. We identify $\rep{80} = \rep{Ad}_9$ and 
$\rep{84} = [\rep{9}]_3$, hence we find their traces can be expressed as 
\equ{
\tr_{\rep{80}} [(iF)^2]_{\SU{9}} = 18\, \tr_{\rep{9}} [(iF)^2]_{\SU{9}},
\qquad 
\tr_{\rep{84}} [(iF)^2]_{\SU{9}} = 21 \, \tr_{\rep{9}} [(iF)^2]_{\SU{9}},
}
in terms of the reference representation $\rep{9}$ of $\SU{9}$. This
confirms that the index of $\E{8}$ is $60$: 
\equ{
\frac 1{60} \tr_{\rep{248}}[(iF)^2]_{\SU{9}} =
\frac 1{60} \Bigl( 18 + 2\cdot 21 \Bigr) 
\tr_{\rep{9}}[(iF)^2]_{\SU{9}}
= \tr_{\rep{9}}[(iF)^2]_{\SU{9}}.
}

\subsubsection*{$\boldsymbol{\SO{14}'}$ quadratic trace}

Using the expression for the character of the spinor representation 
$\SO{2n}$, given in \cite{vanRitbergen:1998pn}, we find for the quadratic 
trace of the spinorial representation $\rep{64}$ of $\SO{14}$ 
\equ{
\arry{c}{
\tr_{\rep{64}} [(iF)^2]_{\SO{14}} = 
2^7\frac{2^2-1}{4 \cdot 2!} B_2\, \tr_{\rep{14}} [(iF)^2]_{\SO{14}} =  
8\, \tr_{\rep{14}} [(iF)^2]_{\SO{14}},
\\[2ex]
\tr_{\rep{91}} [(iF)^2]_{\SO{14}} = 12\, \tr_{\rep{14}} [(iF)^2]_{\SO{14}},
}
}
with the Bernoulli number: $B_2 = 1/6$. For the second relation, we used 
that the adjoint of $\SO{14}$ is obtained as the anti--symmetric
representation $\rep{91} = [\rep{14}]_2$. Following the branching
\eqref{branchings}  of $\E{8}$ to $\SO{14}$ representations gives 
\equ{
\frac 1{60}\tr_{\rep{248}} [(iF)^2]_{\SO{14}} = 
\frac 1{60} \Bigl( 12 + 2 + 2\cdot 8 \Bigr) \tr_{\rep{14}} [(iF)^2]_{\SO{14}}
= \frac 12 \, \tr_{\rep{14}} [(iF)^2]_{\SO{14}}.
}

\subsubsection*{$\boldsymbol{\U{1}'}$ quadratic trace}

Finally, we compute the quadratic traces of the $\U{1}'$ factor in the 
second $\E{8}'$ gauge group. As for the previous $\U{1}$ factor, we
can use the charges given in the branching rules \eqref{branchings}
\equ{
\frac 1{60}\tr_{\rep{248}} [(iF)^2]_{\U{1}'} = 
\frac 1{60} \Bigl( 2 \cdot 14\cdot (\pm 2)^2 
+ 2 \cdot 64 \cdot (\pm 1)^2 \Bigr) [(iF)^2]_{\U{1}'} = 
4\, [(iF)^2]_{\U{1}'}. 
}

\section{Torus wavefunctions with Wilson lines}
\labl{sc:WaveProp}

Here we collect some of the properties of bosonic torus wave functions
that we use in the main text to compute the tadpoles of the gauge
fields. The mode functions $\gf_q(z)$ of the torus are the periodic
scalar functions on $\Cplx^3$ 
\equ{
\left.
\arry{l}{
\gf_q(z + ~~ \hat \imath) = \gf_q(z)
\\[1mm] 
\gf_q(z + \gth\, \hat \imath) = \gf_q(z) 
}\right\}
\quad \Ra \quad 
\gf_q(z) = N_q \, e^{2\pi i (q_i z_i + \bq_\ui \bz_\ui)/R_i},
\quad 
\pmtrx{ q_i \\ \bq_\ui} = 
\frac 1{\bgth - \gth} 
\pmtrx{ ~~\bgth n_i - m_i \\ - \gth n_i + m_i},
\labl{TorusWave}
}
with $n_i, m_i \in \Intr$. The normalization 
$N_q^{-2} = \prod_i R_i^2 \frac {\bgth-\gth}{2i}$  is chosen such that
these wave functions are orthonormal and form a complete set on the
torus $T^6$ 
\equ{
\int_{T^6} \!\d z\,  \gf_q^\dag(z) \gf_{q'}(z) = \gd_{q\,  q'}, 
\qquad 
\sum_{q}  \gf_q(z) \gf_q^\dag(z') = \gd(z - z' - \gG).
\labl{TorusComplete}
}
From these mode functions it is not difficult to obtain algebra valued
mode functions that are periodic up to global gauge transformations. 
\equ{
\arry{l}{
\gf_{q\ga}(z) =  \gf_q(z) \, T(z) T_\ga T\inv(z) 
= \gf_q(z) \, e^{~2\pi i\, a_\ga(z) } \, T_\ga
\\[3mm] 
\gf^{\dag}_{q\ga}(z) =  
\gf^\dag_q(z) \, T(z) T^\dag_\ga T\inv(z) 
= \gf^\dag_q(z) \, e^{\mbox{-}2\pi i\, a_\ga(z) } \, T_\ga^\dag,
\labl{TorusExp}
}
}
using the notation 
\equ{ 
a_\ga(z) = \sum_i\frac 1{R_i} (b_{\ga i} z^i + b_{\ga\ui} \bz^\ui), 
\quad \text{with} \quad 
b_{\ga i} = 
\frac {1 - \bgth}{\gth - \bgth} a_i^I \, w_I(T_\ga), 
\quad 
b_{\ga\ui} = (b_{\ga i})^*. 
\labl{Defb}
} 
These mode functions satisfy the following properties 
\equ{
\arry{lcl}{
\der_i \Bigl( \gf_{q\ga}^\dag(\gth^{-k}z)  \Bigr)  \gf_{q\ga}(z)
& = & 
 - \bgth^{k}\, \gf_{q\ga}^\dag(\gth^{-k}z)\, \der_i \gf_{q\ga}(z), 
\\[2ex]
\der_i \Bigl( \gf_{q\ga}^\dag(\gth^{-k}z) \,\gf_{q\ga}(z) \Bigr) 
& = & 
(1 - \bgth^{k})\, \gf_{q\ga}^\dag(\gth^{-k}z)\, \der_i \gf_{q\ga}(z), 
\\[2ex]
\der_\ui \Bigl( \gf_{q\ga}^\dag(\gth^{-k}z)\, \gf_{q\ga}(z) \Bigr) 
& = & 
(1 - \gth^{k})\, \gf_{q\ga}^\dag(\gth^{-k}z)\, \der_\ui \gf_{q\ga}(z),
\\[2ex]
\gD \Bigl( \gf_{q\ga}^\dag(\gth^{-k}z) \,\gf_{q\ga}(z) \Bigr) 
& = & 
3 \, \gf_{q\ga}^\dag(\gth^{-k}z)\, \gD \gf_{q\ga}(z),
}
\labl{DerProducts}
}
where the Laplacian $\gD = - 2 \sum \der_i \der_\ui$. 
By combining \eqref{TorusComplete} and \eqref{TorusExp}, we obtain the
following identity 
\equ{
\sum_q \gf_{qw}^\dag(\gth^{-k} z) \gf_{qw}(z) = \sum_s 
e^{- 2\gp i\, k s_i a_i^I \, w_I} \frac 1{27} \gd(z - \fZ_s -\gG),
\labl{CombiWaves}
}
where the orbifold delta function is given by \eqref{OrbiDelta}.

\bibliographystyle{paper.bst}
{\small
\bibliography{paper}
}

\end{document}